\documentclass[aps,prd,superscriptaddress,showpacs,twocolumn,nofootinbib,preprintnumbers]{revtex4}
\usepackage{graphicx}
\usepackage{dcolumn}
\usepackage{bm}
\usepackage{natbib}
\usepackage{amsmath}
\usepackage{amssymb}
\usepackage{epstopdf}
\usepackage{hhline}
\usepackage{color}
\usepackage{relsize}
 \usepackage[titletoc,toc,title]{appendix}
\usepackage{array}
\newcolumntype{P}[1]{>{\centering\arraybackslash}p{#1}}
\usepackage{multirow}

\newcommand{\be}{\begin{equation}}
\newcommand{\ee}{\end{equation}}
\newcommand{\bea}{\begin{eqnarray}}
\newcommand{\eea}{\end{eqnarray}}

\begin{document}

\title{General Constraints on Dark Matter Decay from the Cosmic Microwave Background}
\preprint{MIT-CTP/4842}

\author{Tracy R. Slatyer}
\email{tslatyer@mit.edu}
\affiliation{Center for Theoretical Physics, Massachusetts Institute of Technology, Cambridge, MA 02139, USA}

\author{Chih-Liang Wu}
\email{cliang@mit.edu}
\affiliation{Center for Theoretical Physics, Massachusetts Institute of Technology, Cambridge, MA 02139, USA}


\begin{abstract} 
Precise measurements of the temperature and polarization anisotropies of the cosmic microwave background can be used to constrain the annihilation and decay of dark matter. In this work, we demonstrate via principal component analysis that the imprint of dark matter decay on the cosmic microwave background can be approximately parameterized by a single number for any given dark matter model. We develop a simple prescription for computing this model-dependent detectability factor, and demonstrate how this approach can be used to set model-independent bounds on a large class of decaying dark matter scenarios. We repeat our analysis for decay lifetimes shorter than the age of the universe, allowing us to set constraints on metastable species other than the dark matter decaying at early times, and decays that only liberate a tiny fraction of the dark matter mass energy. We set precise bounds and validate our principal component analysis using a Markov Chain Monte Carlo approach and \emph{Planck} 2015 data.
\end{abstract}


\maketitle

\section{Introduction}

	Dark matter (DM) must be stable on timescales comparable to the lifetime of our cosmos -- but it may still decay with a very long lifetime, a subdominant component of the DM might decay on shorter timescales, or decays might transform a slightly-heavier metastable state into the DM we observe today. Such decays are well-motivated and natural in many classes of DM models -- including, for example, R-parity violating decays of the neutralino \cite{BEREZINSKY1991382} or gravitino \cite{Takayama2000388,1126-6708-2007-03-037}, moduli DM \cite{PhysRevD.87.043508}, axinos \cite{Kim200218}, sterile neutrinos \cite{Abazajian:2012ys} and hidden $U(1)$ gauge boson \cite{Chen200971} -- but are unlikely to be probed in any terrestrial experiment, due to their very long timescales. Only indirect searches have the potential to observe DM decay products; furthermore, only studies of the early universe may be able to probe scenarios where a sub-component of DM decays with a lifetime \emph{shorter} than the present age of the universe.

	Energy injection between recombination and reionization will affect the ionization and thermal history of the universe during the cosmic dark ages. Measurements of the cosmic microwave background (CMB) probe the dark ages, and thus provide an avenue to constrain any new physics that would lead to such early energy injections: in particular, the non-gravitational interactions of DM. DM annihilation or decay during and after the epoch of last scattering (z$\sim$1000) will generically inject high-energy particles into the photon-baryon fluid; as these particles cool, they will heat and ionize neutral hydrogen, increasing the residual ionization level after recombination and hence modifying the CMB anisotropy spectrum, changing the gas temperature history, and distorting the black-body spectrum of the CMB \cite{Adams:1998nr,Chen:2003gz,Padmanabhan:2005es,2012MNRAS.419.1294C}. Consequently, accurate measurements of the CMB by recent experiments -- including WMAP, ACT, SPT and \emph{Planck} \cite{2013ApJS..208...19H,Naess:2014wtr,2014ApJ...782...74H,Planck:2015xua} -- can set stringent constraints on the properties of DM. In particular, the impact on the CMB anisotropy spectrum is typically dominated by annihilation or decay at relatively high redshifts, prior to the formation of the first stars, where perturbations to the DM density are small and the astrophysics is simple and well-understood. Consequently, these constraints evade many uncertainties associated with present-day Galactic astrophysics and DM structure formation.
	
	The Standard Model (SM) products of DM annihilation or decay -- which might include gauge bosons, charged leptons, hadrons, or other exotic particles -- will in turn decay to produce spectra of neutrinos, photons, electrons, positrons, protons and antiprotons. Neglecting the contribution of neutrinos, protons and antiprotons (see \cite{Weniger:2013hja} for a discussion of the latter), for precise constraints it is necessary to understand the cooling of photons, electrons and positrons, and their eventual contribution to ionization, excitation and heating of the gas. Early studies \cite{Chen:2003gz, Padmanabhan:2005es} used two simple approximations: (1) that some constant fraction $f$ of the injected energy was promptly absorbed by the gas, with the rest escaping, and (2) that the fraction of absorbed energy proceeding into ionization and excitation is $(1-x_e)/3$, whereas that proceeding into heating is $(1 + 2 x_e)/3$, where $x_e$ is the background hydrogen ionization fraction. Subsequent studies \cite{Slatyer:2009yq,Galli:2013dna} have demonstrated that it is important to account for delayed energy absorption, redshift-dependent absorption efficiency, and the fact that the fraction of deposited energy proceeding into different channels depends on both the redshift and the energy of the primary electron/positron/photon.
	
	A recent analysis \cite{Slatyer:2015kla} has presented interpolation tables describing the power into ionization/excitation/heating from primary electrons, positrons and photons injected at arbitrary redshifts during the cosmic dark ages, with initial energies in the keV $-$ TeV range. This allows easy translation of any model of annihilating or decaying DM into redshift-dependent source functions for excitation, ionization and heating. Using these results, \cite{Slatyer:2015jla} studied the impact on the CMB anisotropy spectrum of keV $-$ TeV photons and $e^+ e^-$ pairs produced by DM annihilation. Extending earlier studies \cite{Finkbeiner:2011dx,Hutsi:2011vx,2012ApJ...752...88F}, that work demonstrated that the imprint on the CMB anisotropy spectrum was essentially identical for \emph{all} models of $s$-wave DM annihilation with keV $-$ TeV annihilation products, up to an overall model-dependent scaling factor, which could be estimated using principal component analysis (PCA). \cite{Slatyer:2015jla} further provided a simple recipe for determining the CMB anisotropy constraints on arbitrary models of annihilating DM: compute the spectrum of electrons, positrons and photons produced by a single annihilation, determine the weighted efficiency factor using the results of \cite{Slatyer:2015jla}, and then apply the bound computed by the \emph{Planck} collaboration on the product of this efficiency factor and the DM annihilation cross section.\footnote{Earlier work  \cite{Finkbeiner:2011dx} applied the same principal component approach to a much broader class of energy injections, with arbitrary redshift dependence, but that work (a) relied on an earlier simplified prescription for the energy deposition, and (b) found that for fully general energy injections, several principal components were needed to adequately describe the impact on the CMB. Restricting ourselves to classes of models that can be described by a single principal component allows for simpler broad constraints.}
	
	In this article, we extend the same approach to the case of decaying DM. Using the public code \texttt{CLASS} \cite{lesgourgues2011cosmic}, we compute the effects on the cosmic microwave background of keV $-$ TeV electrons, positrons and photons injected by DM decay. Scanning over injection energies and species defines a set of basis models, which we use as the input to a PCA. The variance is dominated by the first principal component, which thus largely describes the shape of the perturbation to the CMB anisotropy spectrum from arbitrary DM decays. The coefficients of the basis models in the first principal component trace their approximate relative impact on the CMB, and hence the ``effective detectability'' parameter for photons and $e^+ e^-$ pairs injected at a range of different energies. Once the effective-detectability parameters for both a reference DM-decay model and any other DM-decay model are known, a constraint on the reference model can be approximately translated to all other models. We provide the general recipe and results required to compute effective-detectability parameters for arbitrary models of decaying DM.
	
	We apply the public Markov Chain Monte Carlo (MCMC) code \texttt{Monte Python} \cite{audren7183conservative} to the \emph{Planck} 2015 likelihood to compute the precise limit on our reference model, which we choose (largely arbitrarily) to be DM with a mass of $2\times ( 10^{1.5} + m_e) $ MeV, decaying to $e^+ e^-$ pairs (so the electron and positron each have $\sim 30$ MeV of kinetic energy, which gives rise to the largest signal as we will show later; we will later refer to this reference model loosely as producing 30 MeV $e^+ e^-$). We compute the MCMC limits for several other simple models as a cross-check on our effective-detectability approach, and find good agreement. We provide comparisons of our limits to existing bounds in the literature, finding that our new constraints are stronger than previous bounds for sub-GeV DM decaying primarily to $e^+ e^-$.
	
	While DM must be stable on timescales longer than the age of the universe, a small fraction of the original DM could decay with a much shorter lifetime, or early decays from a slightly-heavier state might liberate a tiny fraction of the DM mass energy. We apply the same PCA approach to decays with lifetimes ranging from $10^{13}$ to $10^{18}$ s; for longer lifetimes, the decays occur after the cosmic dark ages, and the impact on the CMB is indistinguishable from decays with lifetimes longer than the age of the universe.  For shorter lifetimes, the decays occur prior to recombination, and the ionization history is not affected -- we leave studies of the impact on the CMB spectrum for future work (see also \cite{2012MNRAS.419.1294C, Chluba:2013wsa}). We describe the shift of the effective-detectability parameters as a function of the decay lifetime.
		
	In Section \ref{sec:energydeposition}, we summarize our methodology for including the products of DM annihilation and decay in the evolution equations for the gas temperature and ionization level, using the public Boltzmann code \texttt{CLASS}.  In Section \ref{sec:pca}, we briefly review the essentials of PCA, and then proceed to derive the principal components in the CMB anisotropy spectrum induced by DM decay. For our reference model and several other benchmarks, we then compute constraints via a full likelihood analysis of the \emph{Planck} 2015 data, and present results in Section \ref{sec:mcmc}. Finally in Section \ref{sec:discussion} we explain how to apply our results to constrain arbitrary models of DM decay, and present examples and comparisons to previous constraints for various SM final states. We present our conclusions in Section \ref{sec:conclusion}. Supplementary materials, including plots of the higher principal components, and information on supplemental data files which are available at \textbf{http://nebel.rc.fas.harvard.edu/epsilon}, are included in the Appendix. 
	
	Throughout this work, we use the cosmological parameters from \emph{Planck} 2015 data \cite{Planck:2015xua}: $\Omega_b h^2=0.0223$, $\Omega_c=0.1188$, $n_s=0.9667$, ln$10^{10} A_s=3.064$, $\tau=0.066$, and 100$\theta_s=1.04093$.

\section{Energy Injection from Dark Matter}
\label{sec:energydeposition} 

If DM annihilates or decays to SM particles, it will inject energy into the universe at a rate given by, for annihilation and decay respectively:
\begin{align}
\left(\dfrac{dE}{dtdV}\right)^{\text{ann}}_{\text{injected}}=\dfrac{\left<\sigma v \right>}{M_{\chi}}c^2 f_X^2 \Omega^2_{\text{DM}} \rho_c^2 \left(1+z\right)^6, \nonumber \\
\left(\dfrac{dE}{dtdV}\right)^{\text{dec}}_{\text{injected}}=\dfrac{e^{-t/\tau}}{\tau}c^2 f_X \Omega_{\text{DM}} \rho_c \left(1+z\right)^3.
\end{align}
Here $\rho_c$ is the critical density of the Universe in the present day, $\Omega_{DM} \rho_c$ is the present-day cosmological density of cold DM, and $f_X$ is the fraction of the DM (by mass density) that participates in these decay/annihilation processes, evaluated \emph{before} the decays/annihilations have significantly reduced its abundance. $M_{\chi} $ is the DM mass, $\left<\sigma v \right>$ is the thermally averaged cross section for self-annihilating DM, and $\tau$ is the DM decay lifetime. Here we neglect structure formation ; previous studies of the impact of DM annihilation on the CMB anisotropy spectrum have demonstrated that most of the effect arises from high redshifts, $z\sim 600$, where inhomogeneities in the DM density are small \cite{Finkbeiner:2011dx, Poulin:2015pna}. Energy injection from DM annihilations and decays extending until late time, and the possible impact on reionization is studied in \cite{Diamanti:2013bia,Liu:2016cnk,1475-7516-2016-08-054}.

Observable impacts of such injections are controlled by the absorption of this energy by the gas, and the modification to photon backgrounds. The latter effect is generally small for models that inject energy during the cosmic dark ages and are not already excluded \cite{Slatyer:2009yq,Chluba:2013wsa}, so we will focus on constraints arising from the former. We will refer to absorption ``channels'', meaning ionization of hydrogen or helium, excitation or heating of the gas, or distortions to the CMB spectrum.

The amount of energy proceeding into the different absorption channels depends on the energy of the primary injected particle, the redshift of injection, and the background level of ionization at that redshift. Furthermore, injections of energy at some redshift can lead to energy absorption at considerably later times, since the timescale for cooling of photons above a few keV in energy can be comparable to the Hubble time \cite{Chen:2003gz}. Thus computing the energy absorbed in the various channels requires a fully time-dependent treatment of the cooling of the annihilation/decay products, taking into account the expansion of the universe. This has been done in the literature \cite{Slatyer:2015kla}, for keV $-$ TeV photons and $e^+ e^-$ pairs, with results provided as interpolation tables over injection redshift, redshift of absorption and energy of the injected primary particle(s). In general, the CMB signature of an arbitrary model of decaying/annihilating DM is dominated by the effect of photons and $e^+ e^-$ pairs (which may be produced directly in the annihilation/decay, or subsequently by the decay of unstable SM annihilation/decay products). The stable final annihilation/decay products will generally also include neutrinos, protons and antiprotons, but neutrinos can be assumed to escape and the impact of neglecting protons and antiprotons is rather small \cite{Weniger:2013hja}.

Consequently, for any given history of energy injection and spectrum of annihilation/decay products (in the keV $-$ TeV range), these results can be used to compute the energy absorbed into each channel as a function of redshift, as discussed in \cite{Slatyer:2015kla}. It is generally convenient to normalize this quantity to the total energy injected at the same redshift. However, for decays with lifetimes much shorter than the age of the universe, the rate of energy absorption may be non-negligible even after the energy injection from decay has ceased, and so in this case we normalize to the power that \emph{would} be injected without the exponential $e^{-t/\tau}$ suppression. Specifically, given the history of energy absorption into each channel, we define ratio functions $p_{\text{ann/dec},c}(z)$ by:
\begin{align}
\left(\dfrac{dE}{dtdV}\right)^{\text{ann}}_{\text{absorbed},c}=p_{\text{ann},c}(z)c^2 \Omega^2_{\text{DM}} \rho_c^2 \left(1+z\right)^6, \nonumber \\
\left(\dfrac{dE}{dtdV}\right)^{\text{dec}}_{\text{absorbed},c}=p_{\text{dec},c}(z)c^2 \Omega_{\text{DM}} \rho_c \left(1+z\right)^3.
\end{align}
These ratio functions capture both the model-dependent parameters controlling the overall rate and the model-dependent redshift dependence; they completely determine the impact on the CMB. We can also factor out the channel-independent constants to define the channel- and model-dependent efficiency functions $f_c(z)$:
\begin{align}
f_c(z) \equiv \left\{ \begin{array}{cr} p_{\text{ann},c}(z) \left(f_X^2 \dfrac{\left<\sigma v \right>}{M_{\chi}} \right)^{-1} & \, \text{annihilating DM}, \nonumber \\
p_{\text{dec},c}(z) \frac{\tau}{f_X} & \, \text{decaying DM}. \end{array} \right.
\end{align} 
The $f_{c}(z)$ functions for annihilation are thus independent of the overall annihilation rate, and the $f_c(z)$ functions for decay are independent of the decay lifetime if $\tau \gg t$ for all relevant timescales. These definitions are consistent with the definition of $f_c(z)$ employed by \cite{Slatyer:2015kla} for annihilating DM, and also with the definitions of $f(z)$ for annihilating and decaying DM employed by \cite{2013PhRvD..87l3513S}, only now with the efficiency function broken down by absorption channel. The $f_c(z)$ functions are obtained by integrating over the whole past history of energy injection, and depend on both the DM model and whether it is annihilating or decaying (as well as the decay lifetime, if it is not long compared to the age of the universe).

\begin{figure}
 \vspace{2ex}
 \includegraphics[width=4.2cm]{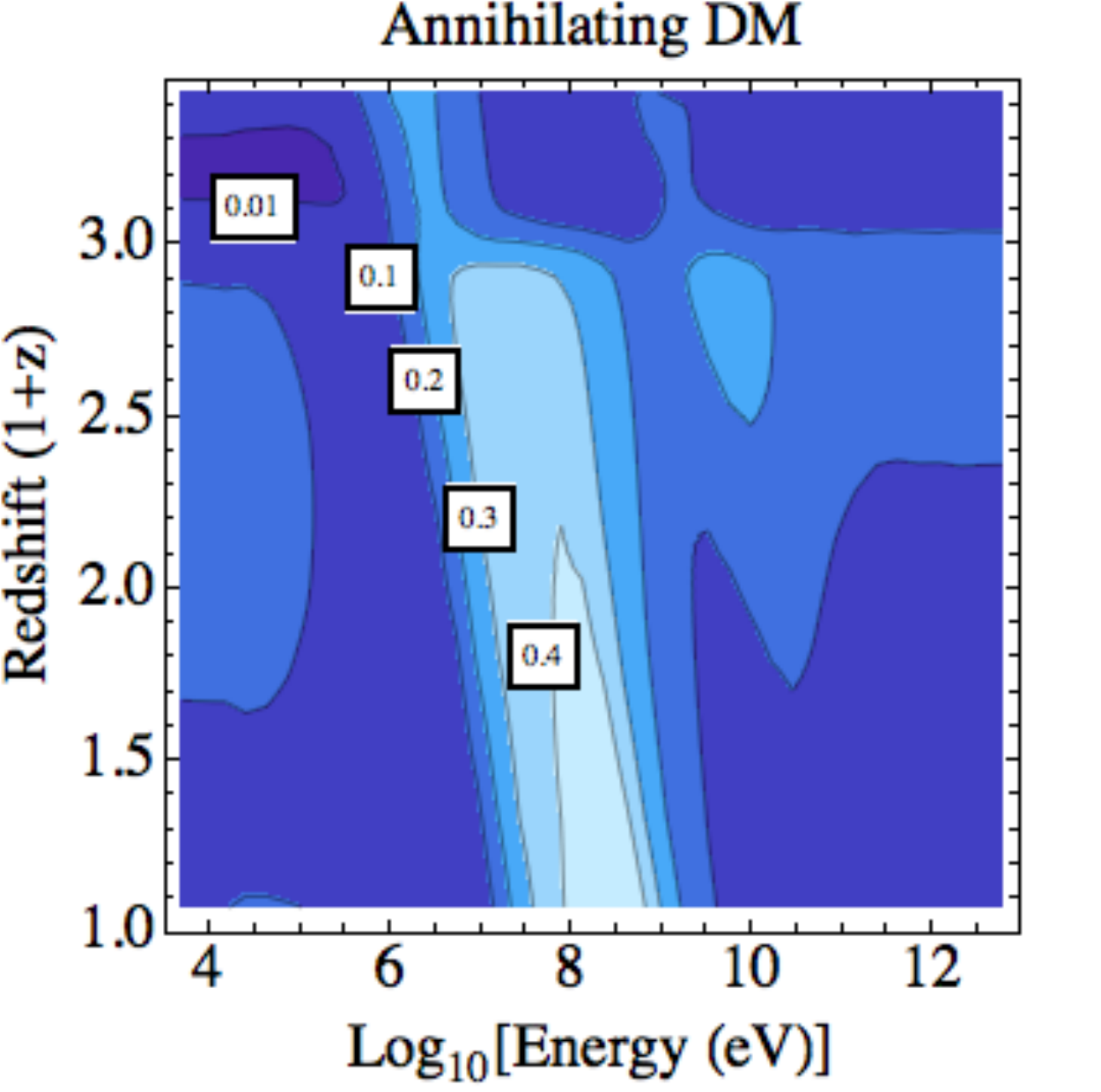}
 \includegraphics[width=4.2cm]{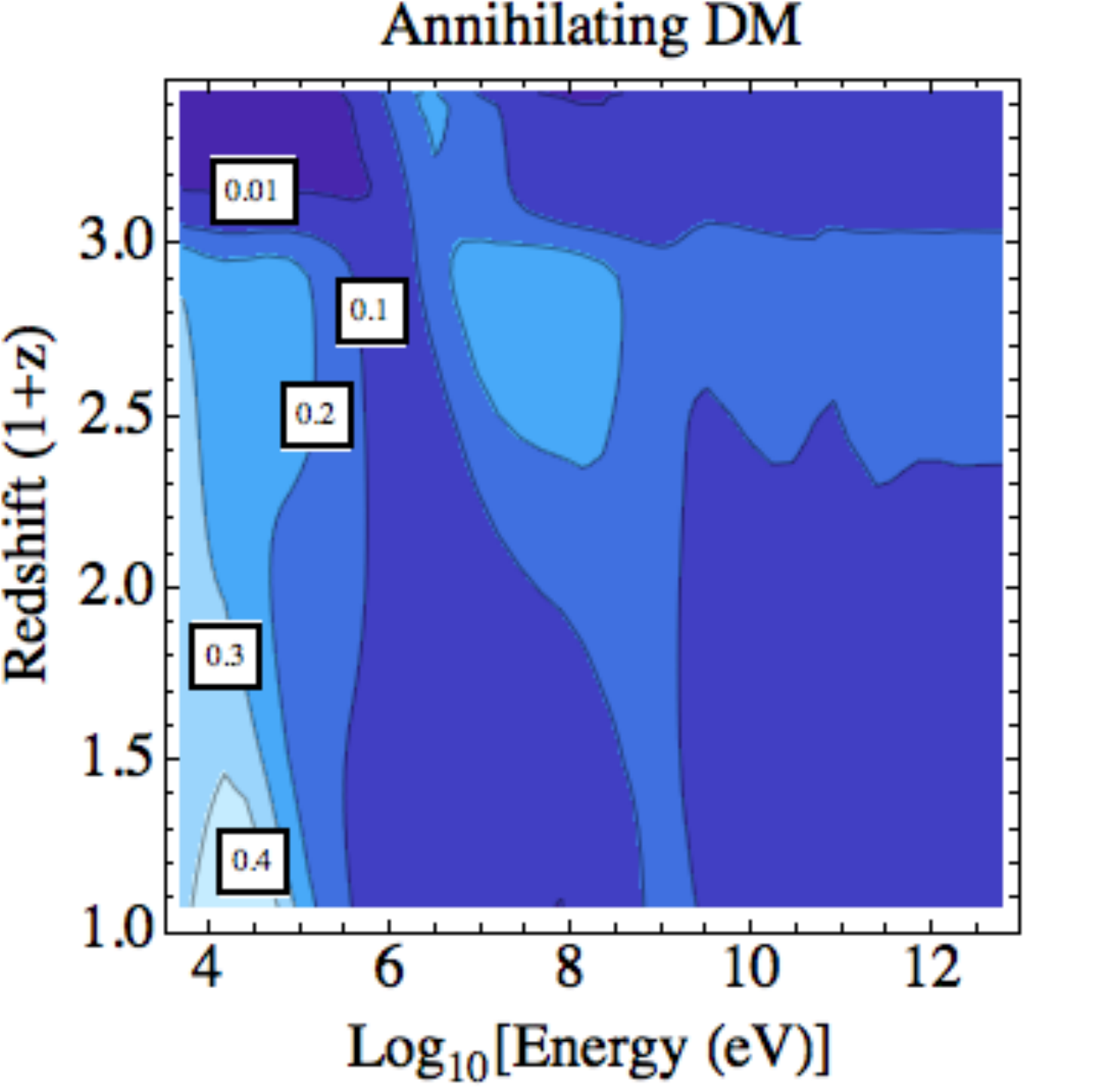} \\[2ex] 
  \includegraphics[width=4.2cm]{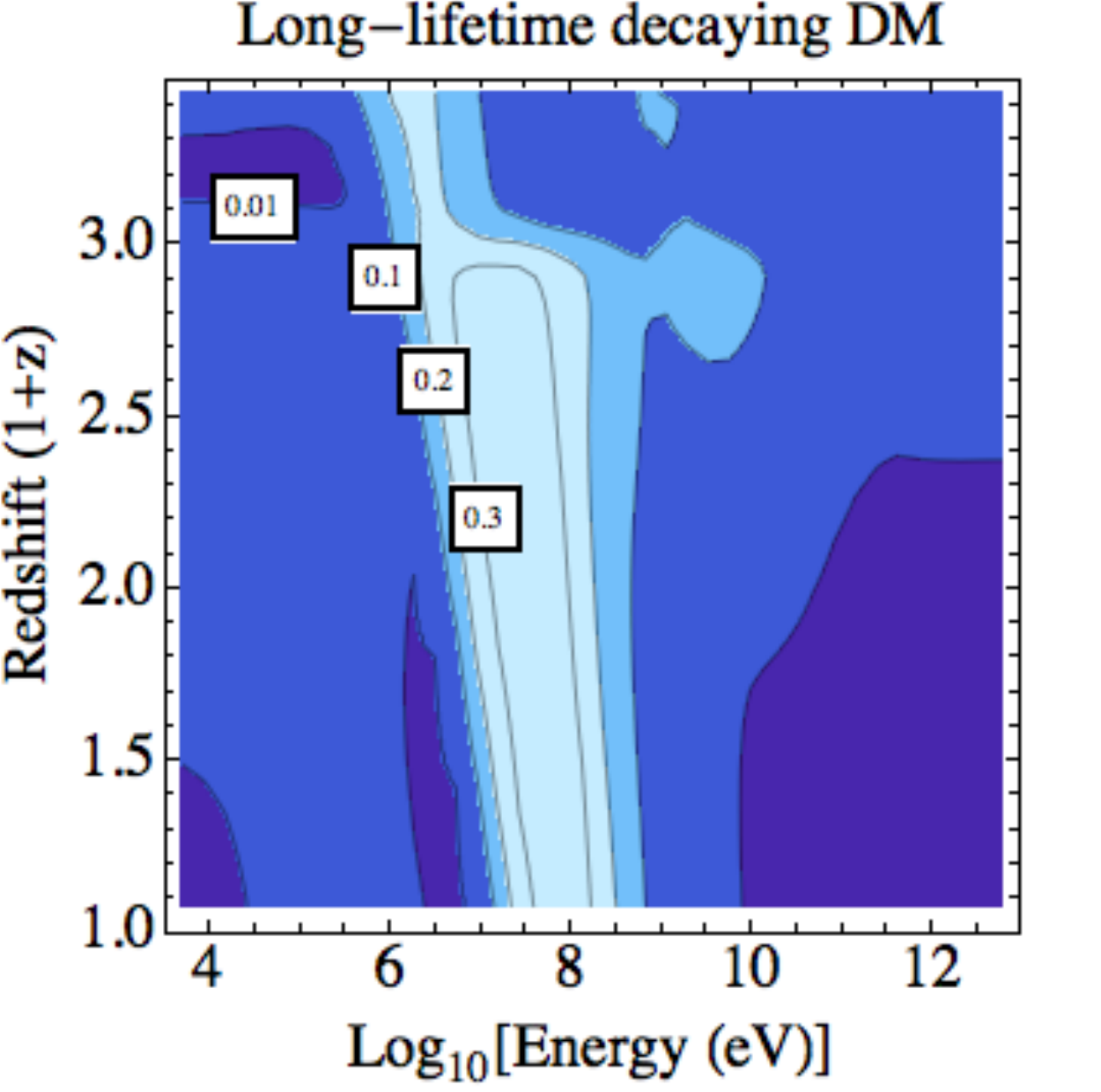}   
  \includegraphics[width=4.2cm]{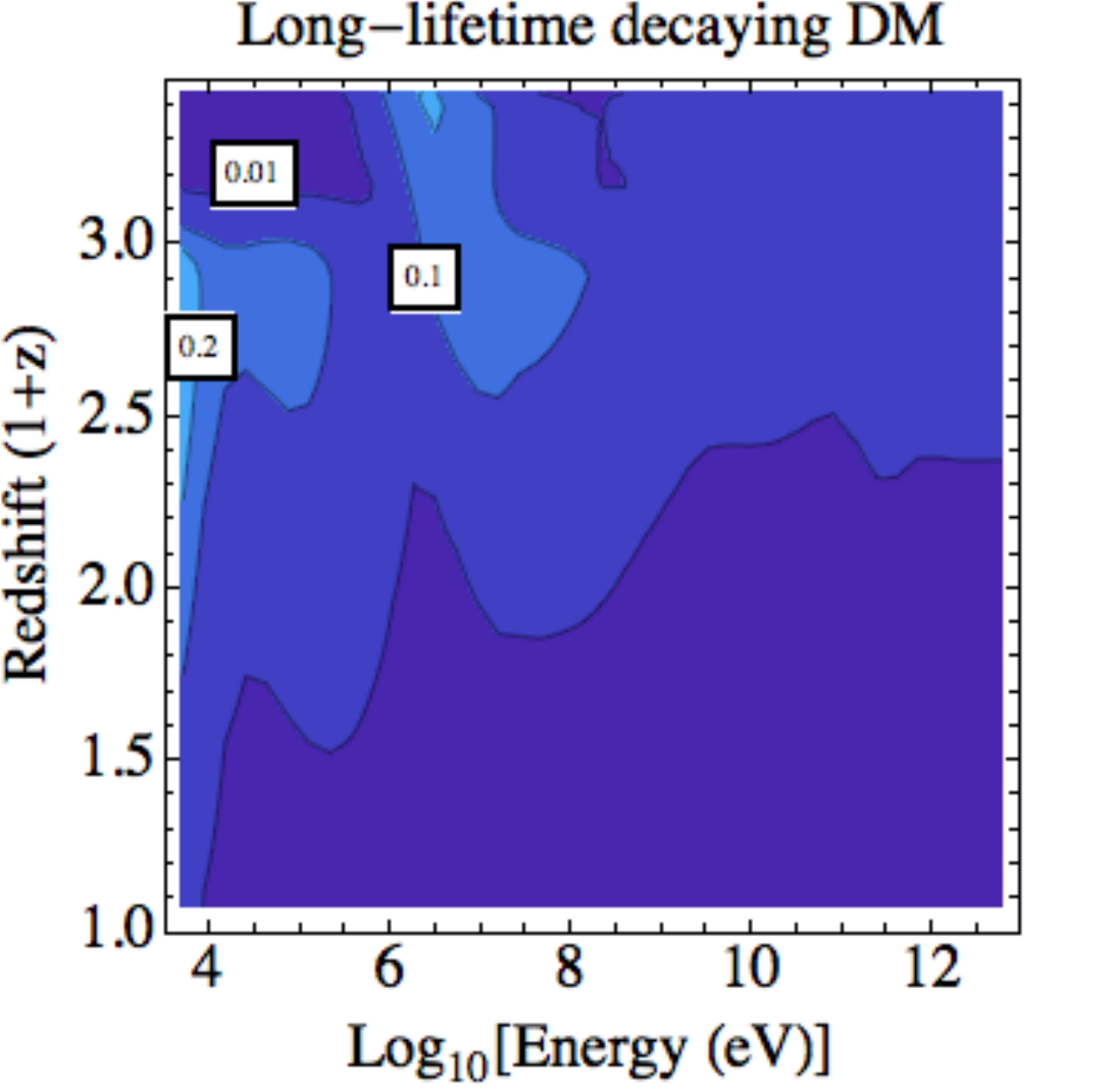} \\[2ex] 
  \includegraphics[width=4.2cm]{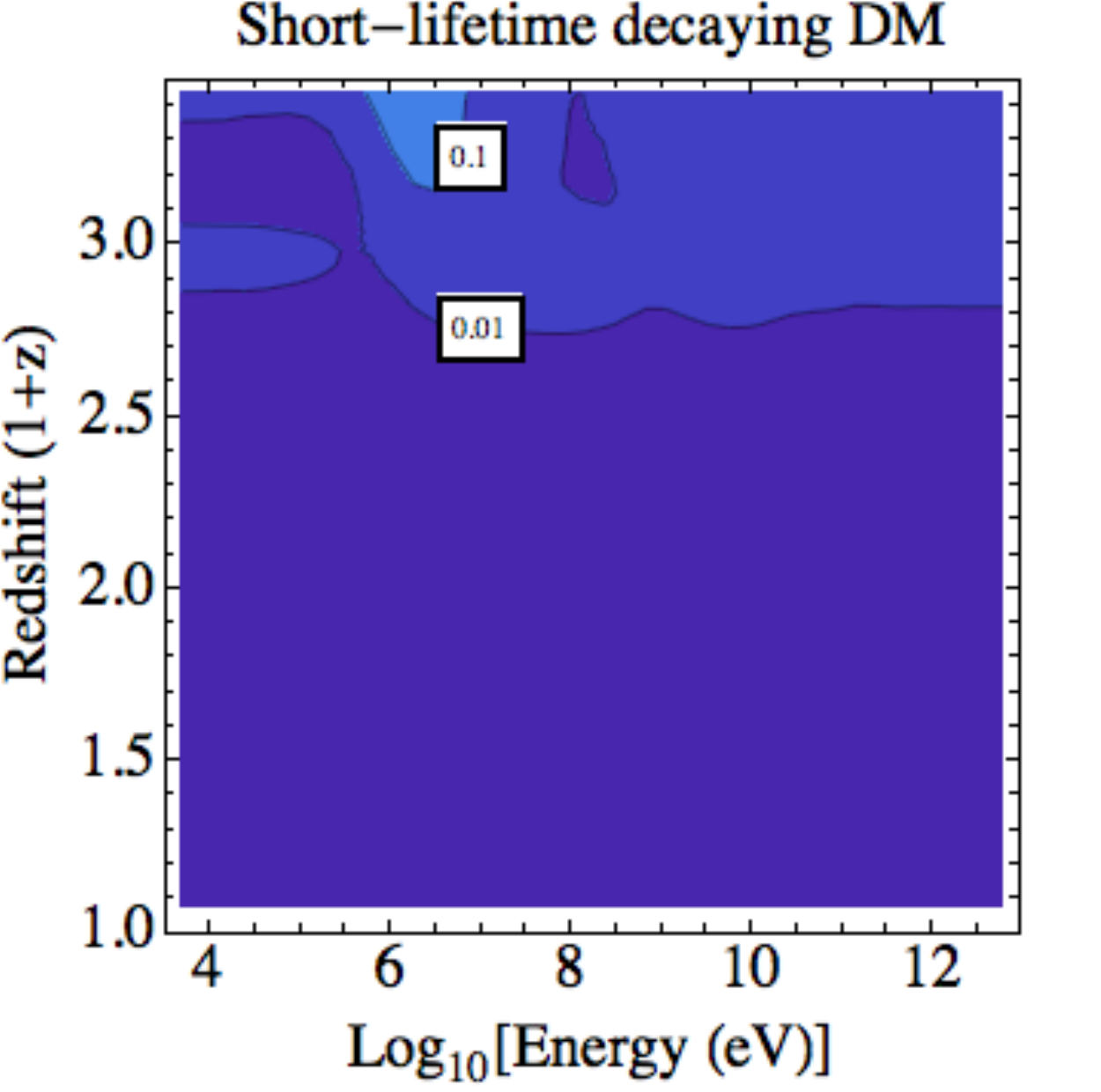} 
  \includegraphics[width=4.2cm]{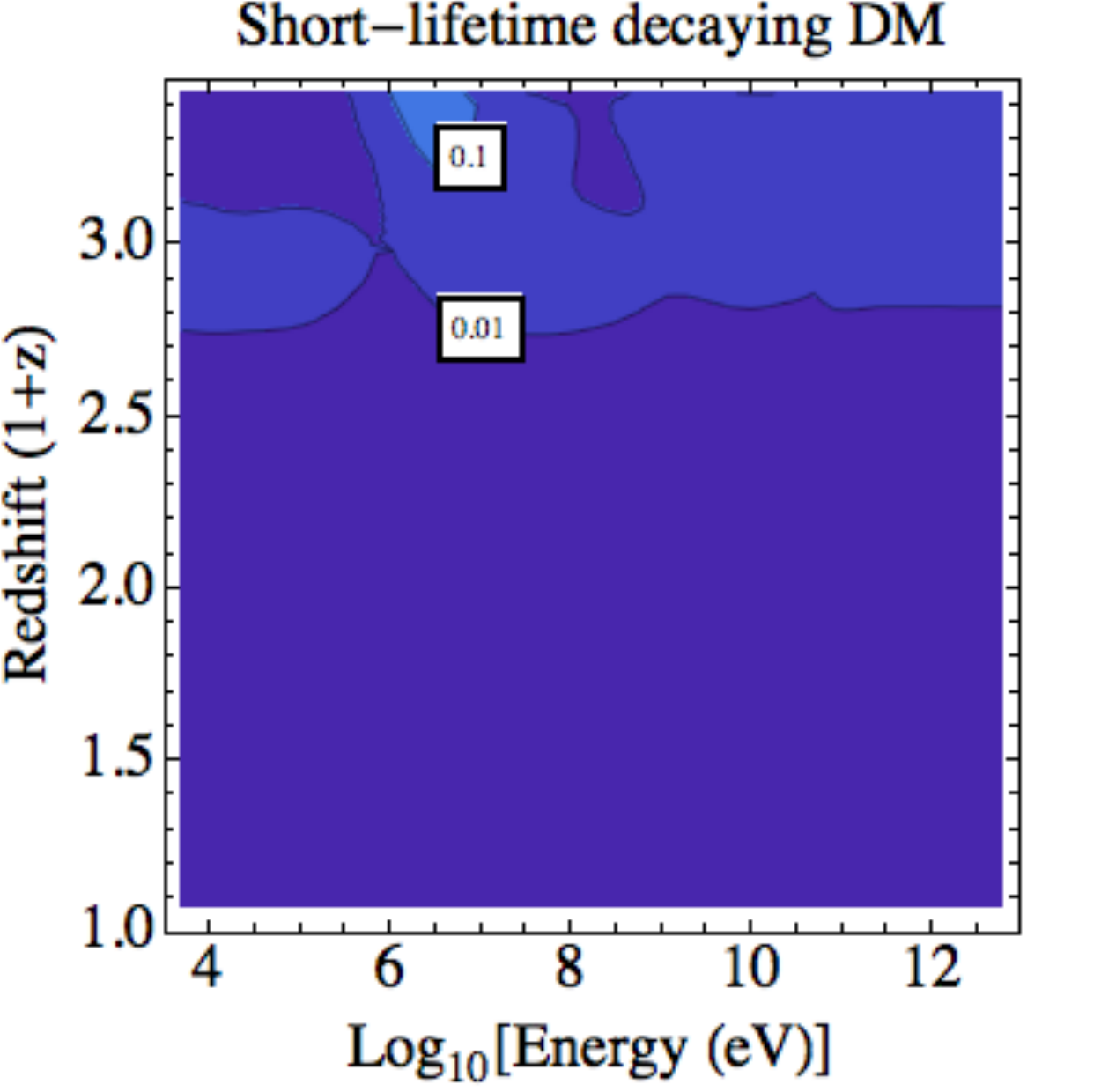}   
 \caption{
Effective efficiency function for energy absorption into hydrogen ionization, for (top) annihilating DM, (middle) long-lifetime ($10^{27}$ seconds) decaying DM , and (bottom) short-lifetime decaying ($10^{13}$ seconds) DM, for initial injection of $e^+e^-$ pairs (left) or photons (right) as a function of redshift-of-deposition and initial (kinetic) energy of one of the injected particles.}
\label{fzplot}
\end{figure}

In Fig. \ref{fzplot} we show the $f_c(z)$ curves for $c=$ ionization on hydrogen, for primary photons and $e^+ e^-$ pairs, as a function of injection energy and redshift of absorption. Different panels show the results for annihilating DM, long-lifetime ($10^{27} $ seconds) decay and short-lifetime ($10^{13} $ seconds) decay.\footnote{A species decaying with such a short lifetime would need to be a subdominant fraction of the DM, or alternatively the decay might only liberate a tiny fraction of its energy.} 

Note the general trend that $f_c(z)$ falls at lower redshifts; this is due to the increased transparency of the universe as it expands, leading to more power escaping into photon backgrounds. The increase in $f_\text{ion}(z)$ for electron/positron energies around $1-100$ MeV is due to the fact that electrons in this energy range upscatter CMB photons to ($\sim 10$ eV $-$ keV) energies where they can efficiently ionize hydrogen; in contrast, for injections of lower-energy $e^+ e^-$ pairs, the upscattered CMB photons are too low-energy to contribute to ionization or excitation (and for sufficiently low energies, the signal becomes dominated by the photons from annihilation of the $e^+ e^-$). For higher electron energies, the upscattered CMB photons are not efficient ionizers and move through a universe that is increasingly transparent to them at low redshifts; consequently, the energy-dependent peak in $f_\text{ion}(z)$ is more pronounced at lower redshifts, as the opacity contrast between sub-keV photons and keV-plus photons becomes more pronounced (see e.g. \cite{Slatyer:2009yq}). The same structure can be seen in $f_\text{ion}(z)$ for injection of photons, at a slightly higher energy; photons in this energy range dominantly lose energy by Compton scattering on electrons, and the resulting energetic electrons go on to produce ionizing photons as discussed above.
	
Following the standard treatment of recombination \cite{Peebles:1968}, we incorporate the power absorbed into the various channels as source terms in the recombination equations, modifying the public \texttt{CLASS} code \cite{lesgourgues2011cosmic}. \texttt{CLASS} has built-in functionality for including DM annihilation, using a simplified prescription for the ratio of power absorbed into different channels; we simply replace this prescription with our more accurate channel-dependent $f_c(z)$ curves.

Specifically, the evolution of the hydrogen ionization fraction $x_e$ (defined as $n_e/n_H$, where $n_H$ is the density of hydrogen and $n_e$ is the density of free electrons) satisfies:
\begin{align}
\dfrac{d x_e}{dz}=\dfrac{1}{\left(1+z\right) H(z)}\left[R_s(z)-I_s(z)-I_X(z)\right],
\end{align} 
where $R_s$ $I_s$ are the standard recombination and ionization rates, and $I_X$ the ionization rate due to DM. This last term has contributions from direct ionization from ground state H atoms, and from the $n = 2$ state:
\begin{align}
I_X(z)=I_{Xi}(z)+I_{X\alpha}(z).
\end{align} 

These contributions can be estimated from the energy absorbed into the hydrogen-ionization (``ion H'') and excitation (``exc'') channels: they correspond to the number of additional ionizations per hydrogen atom per unit time. In terms of the energy absorption rate into these two channels, we can write: 
\begin{eqnarray}
I_{Xi}(z)&=& \left(\dfrac{dE}{dV dt}\right)^{\text{ann/dec}}_{\text{absorbed},\text{ion H}} \frac{1}{n_H(z) E_i} \nonumber \\
I_{X\alpha}(z)&=&\left(1-C\right)\left(\dfrac{dE}{dV dt}\right)^{\text{ann/dec}}_{\text{absorbed},\text{exc}} \frac{1}{n_H(z)E_\alpha}.
\end{eqnarray}    
Here $E_i = 13.6$ eV is the average ionization energy per hydrogen atom, $E_\alpha$ is the difference in binding energy between the 1s and 2p energy levels of a hydrogen atom, and $n_H(z)$ is the number density of hydrogen nuclei. The factor $C$ describes the probability for an electron in the $n = 2$ state to transition to the ground state before being ionized, and is explicitly given by:
 \begin{eqnarray}
C=\dfrac{1+K \Lambda_{2s 1s}n_H\left(1-x_e\right)}{1+K \Lambda_{2s 1s}n_H\left(1-x_e\right)+K\beta_B n_H\left(1-x_e\right)},
\end{eqnarray}  
where $\Lambda_{2s 1s}$ is the decay rate of the metastable $2s$ level, and $K=\lambda_{\alpha}^3/\left(8 \pi H(z)\right)$ accounts for the cosmological redshifting of Lyman-$\alpha$ photons. $H(z)$ is the Hubble factor at redshift $z$, $\lambda_{\alpha}$ is the wavelength of the Lyman-$\alpha$ transition from $2p$ level to $1s$ level, and $\beta_B$ gives the effective photoionization rates for principal quantum numbers $\geq2$.  

Helium ionization follows a similar evolution equation, but we have neglected the effects of energy injection from DM on ionization of helium, as generally the fraction of the injected energy absorbed into helium ionization is small \cite{Slatyer:2015kla}, and the background helium ionization level has little impact on the recombination history \cite{Galli:2013dna}.

A fraction of the energy released by DM goes into heating of the baryonic gas, adding an extra $K_h$ term in the standard evolution equation for the matter temperature $T_b$ (e.g. described in CLASS \cite{lesgourgues2011cosmic}):
\begin{eqnarray}
\left(1+z\right)\dfrac{dT_b}{dz}&=&\dfrac{8\sigma_Ta_RT^4_{CMB}}{3m_ecH(z)}\dfrac{x_e}{1+f_{He}+x_e}\left(T_b-T_{CMB}\right) \nonumber \\
& &-\dfrac{2}{3k_BH(z)}\dfrac{K_h}{1+f_{He}+x_e}+2T_b,
\end{eqnarray}   
with $\sigma_T$ the Thomson cross section, $a_R$ the radiation constant, $m_e$ the electron mass, $c$ the speed of light, and $f_{He}$ the fraction of helium by number of nuclei. The non-standard term is given by
\begin{eqnarray}
K_h= \left(\dfrac{dE}{dV dt}\right)^{\text{ann/dec}}_{\text{absorbed},\text{heat}}\frac{1}{n_H (z)} .
\end{eqnarray}

\section{Principal Component Analysis}
\label{sec:pca}

	Principal component analysis (PCA) provides a systematic approach to deriving broadly model-independent constraints on DM properties. The effects of energy injection from different DM models on the CMB anisotropy spectrum are highly correlated, and consequently can be characterized by a small number of parameters. PCA yields a basis of principal components with orthogonal effects on the CMB anisotropies (after marginalization over the standard cosmological parameters), into which energy injection models can be decomposed; the eigenvalues of these principal components reflect the detectability of their imprint on the CMB. As we will show, for decaying DM with a given lifetime, the first eigenvalue generally dominates the others by roughly an order of magnitude, so that CMB constraints can be estimated with $\mathcal{O}(10\%)$ accuracy by considering only the overlap of a given model with the first principal component.
	
	We follow the general procedure outlined in \cite{Finkbeiner:2011dx}; we refer the reader to that paper for details of our approach. However, for convenience we summarize the key points below.
	
	We are interested in how different energy injections change the anisotropies of the CMB after marginalizing over the standard cosmological parameters. We will characterize our \emph{basis} energy injection models by:
	\begin{itemize}
	\item species (photons or $e^+ e^-$ pairs)
	\item a single energy of injection $E_i$ (in terms of the kinetic energy of one of the injected particles); where relevant, it is assumed that $E_i = M_\chi$ for annihilation to photons, and $E_i = M_\chi - m_e$ for annihilation to $e^+ e^-$ pairs.
	\item redshift dependence of the energy injection profile (annihilation, decay with a lifetime much longer than the age of the universe, or decay with a short enough lifetime to modify the energy injection profile)
	\end{itemize}
	Different basis models are normalized so that $f_X^2 \langle \sigma v \rangle/M_{\chi}$ $\left(f_X/\tau\right)$ is held fixed at some value $p_\text{ref}$ for annihilating (decaying) DM (recall that by $f_{X}$ we mean the fraction of the DM mass density comprised of the decaying/annihilating species, and in the case of short-lifetime decays this fraction is computed \emph{before} a significant fraction of it decays); the effect of each model on the CMB is thus fully characterized by its $f_c(z)$ functions (which are determined by the three factors above). 
	
	In general, we will hold the redshift dependence of the energy injection profile fixed, and then generate $N$ basis models corresponding to different species and energies of injection. (We will perform one analysis where instead we hold the energy of injection and species fixed, and generate basis models corresponding to different energy injection profiles.) Using the modified \texttt{CLASS} code as described in the previous section, we determine the perturbation to the $TT$, $TE$ and $EE$ anisotropy power spectra induced by each basis model, which we denote $(\Delta C_\ell)_i$ for $i=1..N$. The maximum precision mode in \texttt{CLASS} is turned on for this step, so that the calculated power spectrum is stable at the 0.01$\%$ level, and we can probe the impact of very small energy injections. We vary $p_\text{ref}$ and repeat this procedure; this allows us to (a) test the assumption that the $(\Delta C_\ell)_i$ perturbations are linear with respect to the normalization factor $p_\text{ref}$ and (b) determine the derivatives $\partial (\Delta C_\ell)_i/ \partial p_\text{ref}$ in the limit of small $p_\text{ref}$. For each $\ell$ and channel, the derivative is extracted from a polynomial fit, which also allows us to test the extent of nonlinearity. For the standard set of six cosmological parameters, the maximum permitted energy deposition generally lies within the linear regime, although if the energy deposited is too large, the approximation of nonlinearity will eventually break down. For 2$\sigma$ constraints on DM decay lifetimes we will put later, the nonlinearity is within 10 $\%$. 
	
	These derivatives provide us with the transfer matrix components:
	\begin{align}T_{\ell i}=\left\{\dfrac{\partial (\Delta C_\ell^{TT})_i }{\partial p_\mathrm{ref}}, \dfrac{\partial (\Delta C_\ell^{EE})_i}{\partial p_\mathrm{ref}}, \dfrac{\partial (\Delta C_\ell^{TE})_i}{\partial p_\text{ref}}\right\}. \end{align} 
	Here $T_{\ell i}$ labels the components of the $n_\ell \times N$ transfer matrix $T$ mapping generic energy injections (described in the space of basis states) into perturbations to the CMB. Note that each $T_{\ell i}$ is a three-element vector, holding the perturbations to the $TT$, $TE$, and $EE$ anisotropy spectra at that $\ell$.
	
		A generic DM model that annihilates or decays producing particles in the keV $-$ TeV energy range can be approximated as a weighted sum over these basis models (strictly it is an integral; the approximation is one of discretization), in the sense that -- if we assume a linear mapping between energy injections and perturbations to the CMB anisotropy spectrum -- its effect on the CMB will be an appropriately weighted sum of the results for the basis models. Denoting an arbitrary model as $M$ and the basis models as $M_i$, $i=1..N$, we can schematically write $M = \sum_i \alpha_i M_i$; more precisely, by this we mean $(\Delta C_\ell)_M = \sum_i \alpha_i (\Delta C_\ell)_{M_i}$. The $\alpha_i$ coefficients can be trivially determined given the spectrum of annihilation/decay products for $M$ and the DM lifetime or cross section + mass; specifically, 
		\begin{align} \alpha_i \approx  \frac{1}{p_\text{ref}} \left\{ \begin{array}{cr} f_X^2 \frac{\langle \sigma v \rangle}{M_\chi} \frac{E_i \frac{dN_{e^+ e^-,\gamma \gamma}}{d\ln E_i} d \ln E_i}{M_\chi}, &\, \text{annihilating DM} \\  \frac{f_X}{\tau} \frac{2 E_i \frac{dN_{e^+ e^-,\gamma \gamma}}{d\ln E_i} d \ln E_i}{M_\chi}, &\, \text{decaying DM}. \end{array} \right. \label{eq:coeffs} \end{align}
		
		Here $ \frac{dN_{e^+ e^-,\gamma \gamma}}{d\ln E_i}$ describes the spectrum of $e^+ e^-$ or $\gamma \gamma$ pairs at $E_i$ per annihilation/decay\footnote{Note that one could also write $ \frac{dN_{\gamma \gamma}}{d\ln E_i} = \frac{1}{2} \frac{dN_{\gamma}}{d\ln E_i}$ (the photon spectrum) and $ \frac{dN_{e^+ e^-}}{d\ln E_i} = \frac{dN_{e^+}}{d\ln E_i}$ (the positron spectrum), assuming charge symmetry.}  (i.e. each member of the pair has kinetic energy $E_i$), and $d \ln E_i$ describes the spacing between the sample energies, which should be chosen to cover the whole spectra of photon and $e^+ e^-$ pairs (this is the discretization approximation), such that,
		\begin{align} & \sum_i 2 E_i \frac{dN_{e^+ e^-,\gamma \gamma}}{d\ln E_i} d \ln E_i \nonumber \\
		& \approx \int 2 E \frac{dN_{e^+ e^-}}{d\ln E} d\ln E + \int 2 E \frac{dN_{\gamma \gamma}}{d\ln E} d\ln E, \end{align} 
		which gives the total energy in photons, electrons and positrons per annihilation/decay. 
		
		For DM that both annihilates and decays with a long lifetime, the two contributions to energy injection can simply be added; in our formalism they will generally contribute to different basis models, characterized by different redshift-dependences for the energy injection history. (For DM that annihilates and decays with a \emph{short} lifetime, the redshift dependence of the annihilation will be different to that assumed here, and require a separate analysis.)
	
	The perturbation to the CMB anisotropy due to a general model is then given by $ (\Delta C_\ell)_M \approx \sum_i \alpha_i p_\text{ref} T_{\ell i} = p_\text{ref} T \cdot \vec{\alpha}$, where $\vec{\alpha}$ holds the model coefficients describing its overall normalization and spectrum.
		
Using the transfer matrix, we can construct the $N\times N$ Fisher matrix $F_e$ as
\begin{align}
\left(F_e\right)_{ij}=\sum_{\ell}T_{\ell i}^{T}\cdot  \mathsmaller{\mathsmaller{\sum\nolimits_{\text{cov}}^{-1}}} \cdot \mathsmaller{T_{\ell j}},
\end{align} 
where $\mathsmaller{\sum\nolimits_{\text{cov}}}$ is the appropriate covariance matrix for the anisotropy spectra
\begin{align}
\mathsmaller{\sum\nolimits_{\text{cov}}} & = \dfrac{2}{2\ell+1}\times \\
&\left(
\begin{matrix}
  \left(C^{TT}_ {\ell}\right)^2 &  \left(C^{TE}_ {\ell}\right)^2 & C^{TT}_ {\ell}C^{TE}_ {\ell} \\
  \left(C^{TE}_ {\ell}\right)^2 & \left(C^{EE}_ {\ell}\right)^2 & C^{EE}_ {\ell}C^{TE}_ {\ell} \\
  C^{TT}_ {\ell}C^{TE}_ {\ell} & C^{EE}_ {\ell}C^{TE}_ {\ell}  &  \left(C^{TE}_ {\ell}\right)^2 + C^{TT}_ {\ell}C^{EE}_ {\ell} 
 \end{matrix}
\right).
\end{align} 

	For experiments that are not cosmic variance limited (CVL), we need to include the effective noise power spectrum. In this work, we use the same noise spectrum as in \cite{Finkbeiner:2011dx}. We consider WMAP7, \emph{Planck} and an experiment that is CVL up to $\ell = 2500$ for all the anisotropy spectra we consider (previous studies have indicated that the effect on the CMB is largest at low to intermediate $\ell$ values \cite{Padmanabhan:2005es}). The effect of partial sky coverage is included by dividing $\mathsmaller{\sum\nolimits_{\ell}}$ by $f_{sky} =  0.65$. The diagonal elements $\left(F_e\right)_{ii}$ describe the (squared) signal significance per $p_\text{ref}$ for basis model $M_i$, before marginalization over the existing cosmological parameters. 
	
	It is critical to marginalize over the standard cosmological parameters, as they can have non-negligible degeneracies with energy injections \cite{Finkbeiner:2011dx}. We use \texttt{CLASS} to study the impact of small variations of the cosmological parameters and to construct the transfer matrix from variations in those parameters to changes in the CMB anisotropies. The full marginalized Fisher matrix can be constructed as:
\begin{align}
F_0=\left(\begin{matrix}
  F_e & F_v  \\
  F_v^T & F_e  
 \end{matrix}\right)
\end{align} 
where $F_e$ is the pre-marginalization Fisher matrix, $F_c$ is the Fisher matrix for the cosmological parameters, and $F_v$ describes the cross terms. The usual prescription for marginalization is to invert the Fisher matrix, remove the rows and columns corresponding to the cosmological parameters, but when the number of energy deposition parameters is much greater than the number of cosmological parameters, is is helpful to take advantage of the block-matrix inversion and write the marginalized Fisher matrix as $F=F_e-F_vF_c^{-1}F_v^T$.

Diagonalizing the marginalized Fisher matrix $F$,
\begin{align}
F=W^T\Lambda W, \Lambda=\text{diag}\left(\lambda_1,\lambda_2,....,\lambda_N\right)
\end{align}
we obtain a basis of $N$ eigenvectors / principal components $\vec{e}_i$, $i=1..N$, by reading off the rows of $W$, with corresponding eigenvalues $\lambda_i$, $i=1..N$. These principal components lie in the space of coefficients of the basis models, i.e. they correspond to a set of coefficients of basis models; while the normalization of the principal components is rather arbitrary, we will choose to normalize them so that in this space they are orthonormal vectors. We rank the principal components by eigenvalue, such that $\vec{e}_1$ has the largest eigenvalue.
	
	In general, we can then determine the impact of an arbitrary model on the $C_\ell$'s, orthogonal to the standard cosmological parameters, by simply taking the dot product of its coefficients $\{\alpha_i\}$ with the first PC. Where the first eigenvalue dominates the variance (i.e. is large compared to the sum of all other eigenvalues), it can be thought of as a weighting function, describing the effect of energy injection on the CMB as a function of different injection species and energies.

To be explicit, let us write an arbitrary model of decay/annihilation as $M = \sum_{i=1}^N \alpha_i M_i(z)$ as above, and $\vec{\alpha} = \{\alpha_1, \alpha_2, ... \alpha_N\}$. In the Fisher-matrix approximation (which assumes linearity and a Gaussian likelihood), we can estimate the $\Delta \chi^2$ for model $M$ relative to the null hypothesis of no energy deposition as $\Delta \chi^2 = \sum_{i=1}^{N_\text{max}} (\vec{\alpha} \cdot \vec{e}_i)^2 p_\text{ref}^2 \lambda_i$ (since the eigenvalues of the Fisher matrix describe (significance per $p_\text{ref}$)$^2$, and hence have units $1/p_\text{ref}^2$), where $N_\text{max}$ is the number of principal components we choose to include. From this, we can forecast constraints on decay lifetime; for example, the 2$\sigma$ limit corresponds approximately to the constraint:
\begin{align}
p_\text{ref} < \dfrac{2}{\sqrt{\sum_{i=1}^{N_\text{max}} \lambda_i (\vec{\alpha} \cdot \vec{e}_i)^2}}.\label{eq:constraint}
\end{align}
In particular, for the basis models $M_k$, where $\alpha_j= \delta_{jk}$, we can estimate the constraint on the normalization parameter (which recall is defined to be $f_X^2 \langle \sigma v \rangle/M_\chi$ for annihilation, or $f_X / \tau$ for decay) to be:
\begin{align}
p_\text{ref} < \dfrac{2}{\sqrt{\sum_{i=1}^{N_\text{max}} \lambda_i (\vec{e}_i)_k^2}} \lesssim \dfrac{2}{\sqrt{ \lambda_1}} \times \frac{1}{(\vec{e}_1)_k}.
\end{align}
We see that when the first PC dominates, its component in the direction of a given basis model is inversely proportional to the constraint on $p_\text{ref}$ for that model, and thus directly proportional to the constraint on the decay lifetime $\tau$, for fixed decaying fraction $f_X$. For annihilating DM, the PC components are inversely proportional to the constraints on $\langle \sigma v \rangle/M_\chi$.

\begin{figure}
    \includegraphics[width=6.5cm]{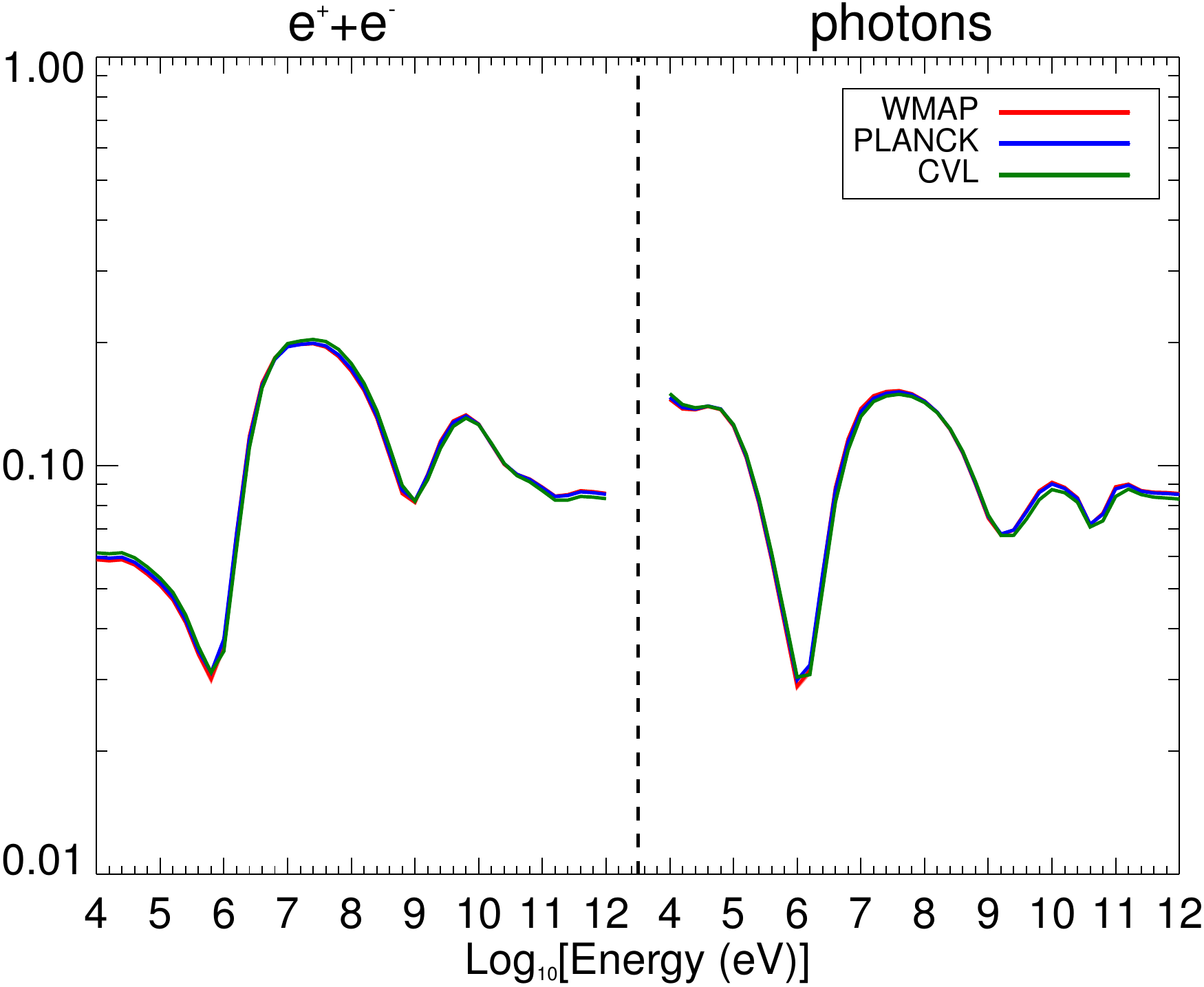}\\[1ex]
   \includegraphics[width=6.5cm]{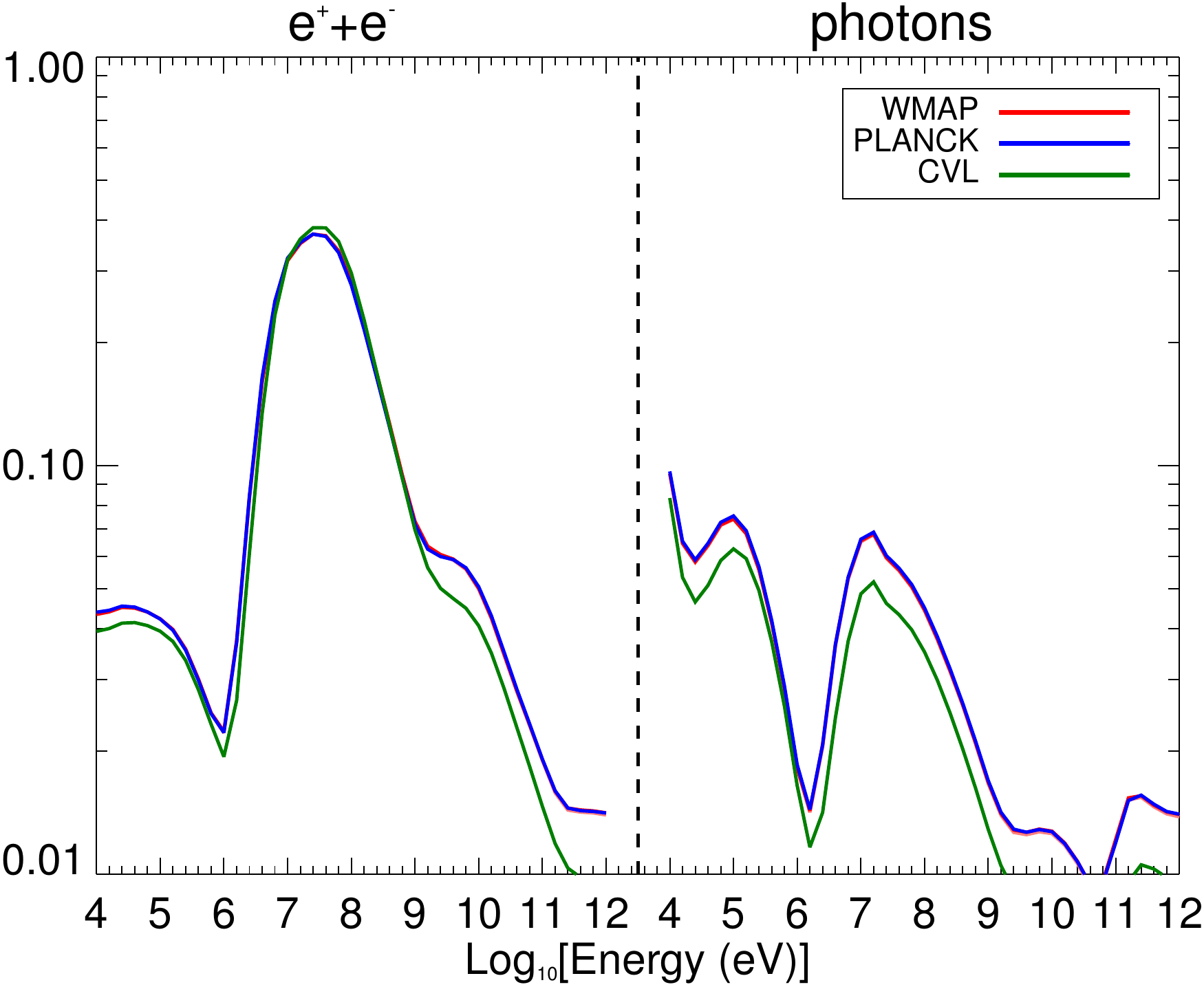}\\[1ex]
   \includegraphics[width=6.5cm]{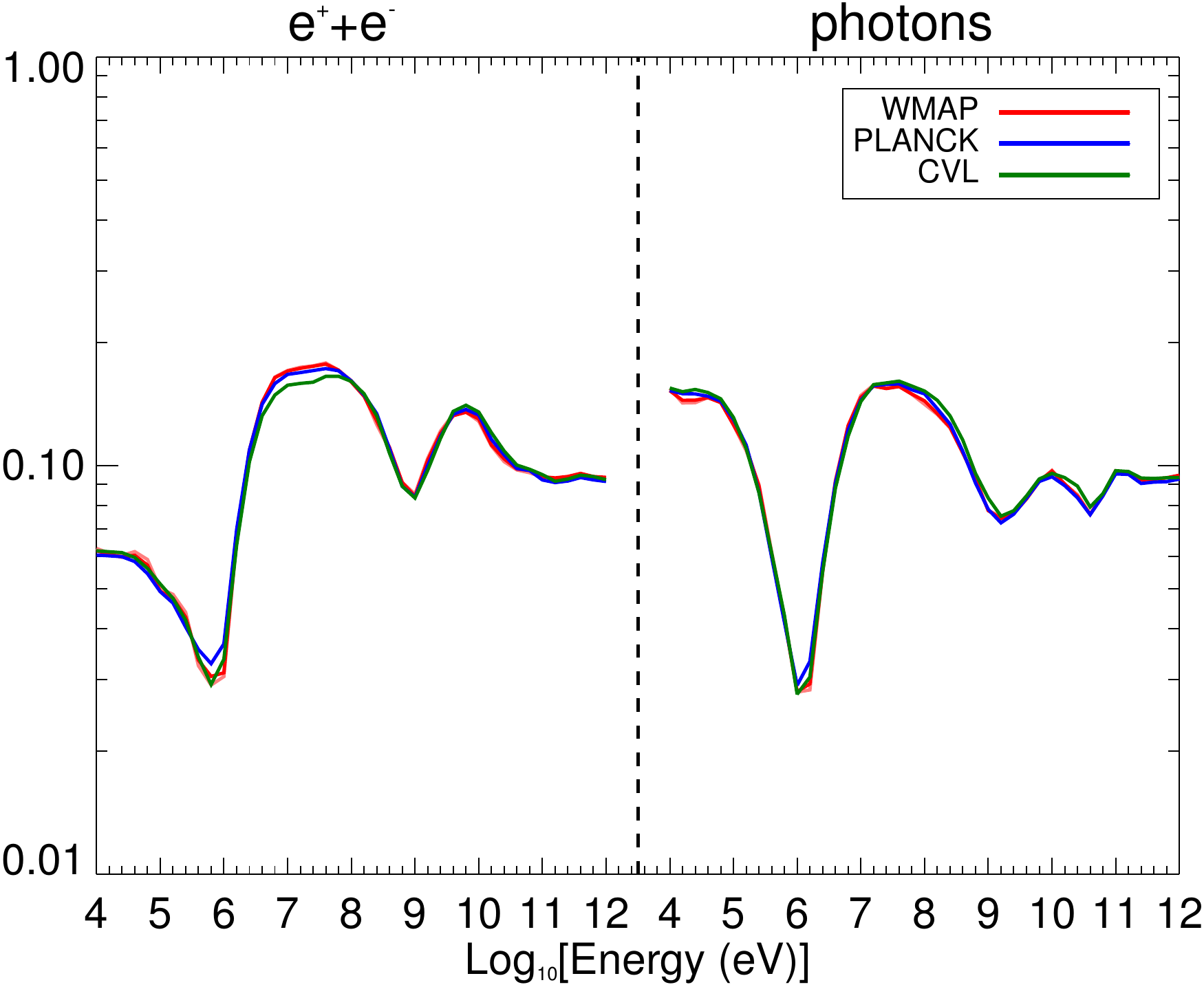}\\[1ex]       
  \caption{
   The first principal components for WMAP7, \emph{Planck} and a CVL experiment, after marginalization over the cosmological parameters, for (top) annihilating DM, (middle) long-lifetime ($10^{27}$ seconds) decaying DM, and (bottom) short-lifetime ($10^{13}$ seconds) decaying DM . The $x$-axis describes the injection energy (kinetic energy for a single particle) for $e^+ e^-$ pairs and photons.}
\label{PCAfirst}
\end{figure}

In Fig. \ref{PCAfirst} we show the first PC (after marginalization) for annihilating, long-lifetime, and (one example of) short-lifetime decaying DM, with a lifetime of $\tau = 10^{13} s$. Here we have labeled the various basis models $M_i$ by their energy-of-injection and species.

The largest eigenvalue, corresponding to the first PC, accounts for 97.0\% of the variance for long-lifetime decay, more than 99.9\% of the variance for annihilation, and 95.7\% of the variance for an example of short-lifetime decay ($\tau = 10^{13} s$). Thus in these three cases the first PC generically dominates the constraints, and we expect restricting ourselves to the first PC to give results accurate at the level of $\mathcal{O}(10\%)$. The approximation of dropping later PCs is much better for the annihilation case, where it is unlikely to induce even percent-level error.

In this case, therefore, the curves in Fig. \ref{PCAfirst} directly map to the strength of the constraint that can be set on $p_\text{ref}$ by the CMB, or equivalently, the degree to which injection of particles with a given energy/species will dominate any signal in the CMB. The PC for annihilation closely matches the equivalent $f_\text{eff}$ curve presented in \cite{Slatyer:2015jla} (up to an irrelevant normalization factor). We see that while for annihilating DM no single energy dominates the signal, in the decay case the first PC is peaked around injection of 30 MeV $e^+ e^-$ pairs for long lifetimes. As mentioned earlier, we attribute this peak to the high efficiency of ionization by the secondary products of $\sim 10-100$ MeV $e^+ e^-$ pairs, and its increased dominance in the case of decaying DM to the fact that the universe is more transparent at the lower redshifts where the signal from decaying DM is peaked (compared to the higher redshifts that provide most of the signal for annihilating DM models). 

\begin{figure}
    \includegraphics[width=6.5cm]{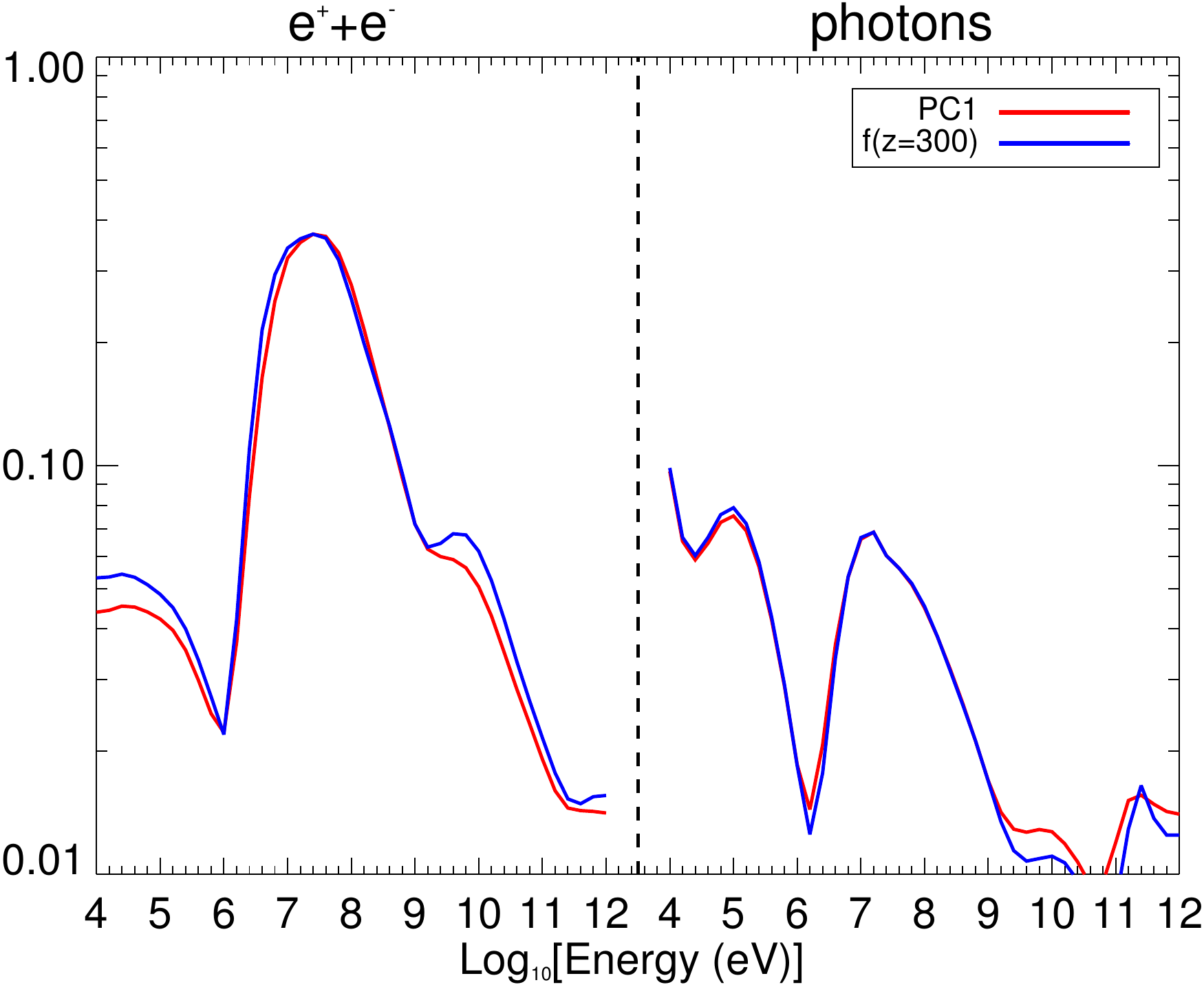}  
  \caption{
   Comparison between the first principal component for \emph{Planck} (red line), as described in Fig.~\ref{PCAfirst}, and $f_c(z=300)$ for hydrogen ionization (blue line), as described in Fig.~\ref{fzplot}.}
\label{fzpc1}
\end{figure}

To confirm our physical understanding of this peak, note that from the general analysis in \cite{Finkbeiner:2011dx}, we expect the impact of DM decay on the CMB to be dominated by redshifts around $z\sim 300$. In Fig.~\ref{fzpc1} we compare the behavior of the first PC to the $f_c(z)$ curve for hydrogen ionization from DM decay at $z=300$, for the photon/electron/positron energies of our basis models (i.e. a horizontal slice through the middle row of Fig.~\ref{fzplot}). We see that the agreement is excellent. Similarly, the $f_\text{eff}$ curve for annihilation \cite{Slatyer:2015jla} is closely approximated by $f_c(z)$ for hydrogen ionization evaluated at $z=600$.

We compute the expected constraint on the decay lifetime, assuming long-lifetime decay, using the first 1-2 PCs; the results are shown in Table  \ref{tab:pcforecast} for a range of injection energies and species (energies refer to kinetic energies). We can see that including the second PC changes the constraints by less than 10 $\%$ in most cases, although it can be a larger effect ($\mathcal{O}(30\%)$) where the overlap with the first PC is small.  This principally occurs for heavier DM; as we will discuss in the next section, these constraints are most interesting for MeV $-$ GeV DM. Contributions from higher PCs are negligible.

\begin{table}
\def\arraystretch{1.3}
\begin{tabular}{ c|P{2.0cm}|m{2cm} c}
\hline
species & energies  & PC1 & PC1+PC2    \\  
 \hline 
 \multirow{5}{*}{electron} &
 10keV   & $0.36$  & $0.37$   \\
 & 1MeV  &   $0.19$  &  $0.19$  \\
 & 100MeV  & $2.49$ & $2.54$ \\
 & 10GeV  &   $0.42$ & $0.45$  \\
 & 1TeV   & $0.11$  & $0.13$ \\   \hline
 \multirow{5}{*}{photon} &
 10keV  & $0.81$  & $0.84$   \\
 & 1MeV  &$0.15$  & $0.16$ \\
 & 100MeV  &$0.37$ & $0.41$ \\
 & 10GeV     &$0.10$  & $0.13$  \\
 & 1TeV  &$0.11$ & $0.14$ \\
 \hline
\end{tabular}
\caption{Forecast \emph{Planck} lower bounds on decay lifetime in units of $10^{25}$ s, at 95 $\%$ confidence, using PCA, for decays to $e^+ e^-$ pairs and photons at a range of energies. Here ``electron'' always labels the electron in an $e^+ e^-$ pair. The first column shows the forecast using only the first principal component, the second the forecast including the first two principal components.}
\label{tab:pcforecast}
\end{table}

\begin{figure}
  \includegraphics[width=6.5cm]{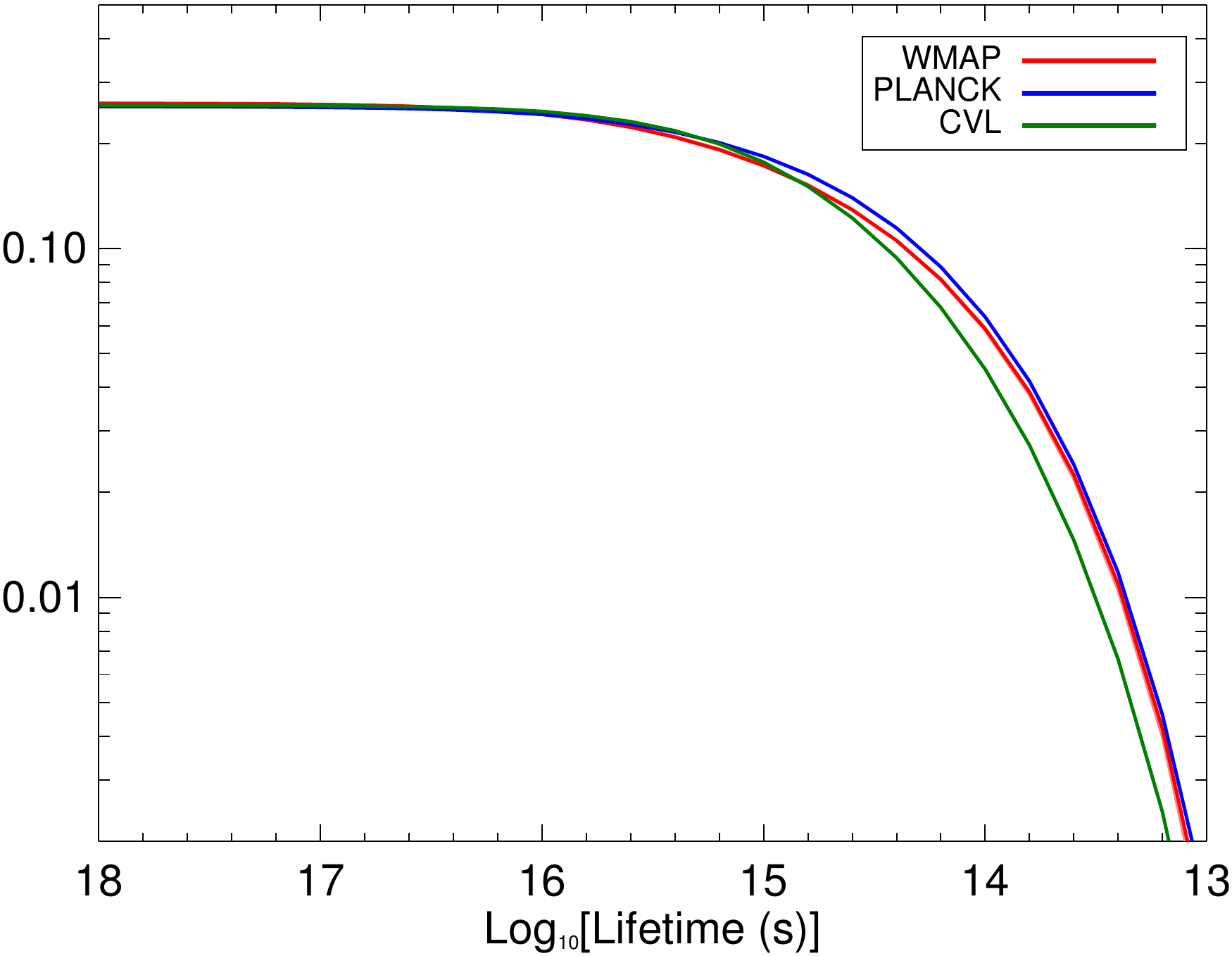}      
  \caption{
   The first principal components for WMAP7, \emph{Planck} and a CVL experiment, after marginalization over the cosmological parameters, for a set of basis models corresponding to energy injection of $e^+ e^-$ pairs with injection energy 30 MeV, with varying decay lifetimes.}
\label{lifetime}
\end{figure}

	As mentioned above, we can also choose our basis models to represent decaying DM with different lifetimes, but with fixed injection energy and species (and as previously, fixed $p_\text{ref}$). Since the strongest CMB signal comes from around 30 MeV (in kinetic energy per particle) electron-positron pairs in the case of decaying DM with a long lifetime, we fix the injection energy to this value, consider only $e^+ e^-$ pairs, and now vary the decay lifetime between basis models. Repeating the PCA described above, we find that in this case the eigenvalue of the first principal component is 98.0\% of the total variance, again dominating the later principal components. This first principal component is shown in Fig.~\ref{lifetime}, where now we have labeled the basis models by their decay lifetimes. 
	
	 As previously, Fig.~\ref{lifetime} can be understood as displaying the estimated relative strength of constraints from the CMB on decaying DM with different lifetimes, assuming the annihilation products are 30 MeV photons or electron-positron pairs. We see that sufficiently short-lifetime DM is almost irrelevant to the constraints; this is expected, since decays occurring before recombination have very little impact on the ionization history. Precise CMB constraints on such short-lifetime decays are difficult to obtain, as if we raise $f_X$ to the point where signals from late redshifts can be measured (above numerical error) in \texttt{CLASS}, there is a very large energy injection in the early universe, and linearity certainly breaks down. Thus this case would require a full likelihood analysis; however, decays with lifetimes less than $\sim 10^{13}$ s are likely to be more tightly constrained by probes of the universe's earlier history, e.g. big bang nucleosynthesis.

\section{Constraints from \emph{Planck} 2015 Data}
\label{sec:mcmc}

	The forecast constraints we have calculated so far are  limited by the assumptions of a Gaussian likelihood and linearity, which are inherent to the Fisher matrix approach. To go beyond these assumptions and find directly the posterior distributions of the cosmological parameters, including the DM decay lifetime, we use the publicly available Markov Chain Monte Carlo (MCMC) parameter estimation code \texttt{Monte Python}, interfaced with \texttt{CLASS}. For the inference procedure, we use the \emph{Planck} 2015 data including three likelihoods; (i) the low-$\ell$ temperature and LFI polarization (bflike, $2 \leq\ell\leq29$), (ii) the high-l plike TTTEEE ($30 \leq\ell\leq2058$) likelihood, and (iii) the lensing power spectrum reconstruction likelihood. 

	We perform the analysis assuming flat priors on the following six cosmological parameters $\omega_b$, $\omega_c$, $n_s$, ln$10^{10} A_s$, $\tau$, and 100$\theta_s$, as well as a new parameter "decay", given by the inverse of the DM decay lifetime in units of s$^{-1}$. Our treatment of the energy deposition is the same as described in the previous sections.  We adopt the Gelman-Rubin convergence criterion (variance of chain means divided by the mean of the chain variances), ensuring that the corresponding R $-$ 1 fell below 0.01.  Our constraints and the 1-D and 2-D likelihood contour plots are obtained after marginalization over the remaining standard nuisance parameters in the \texttt{Monte Python} package. 

\begin{figure}
    \includegraphics[width=8cm]{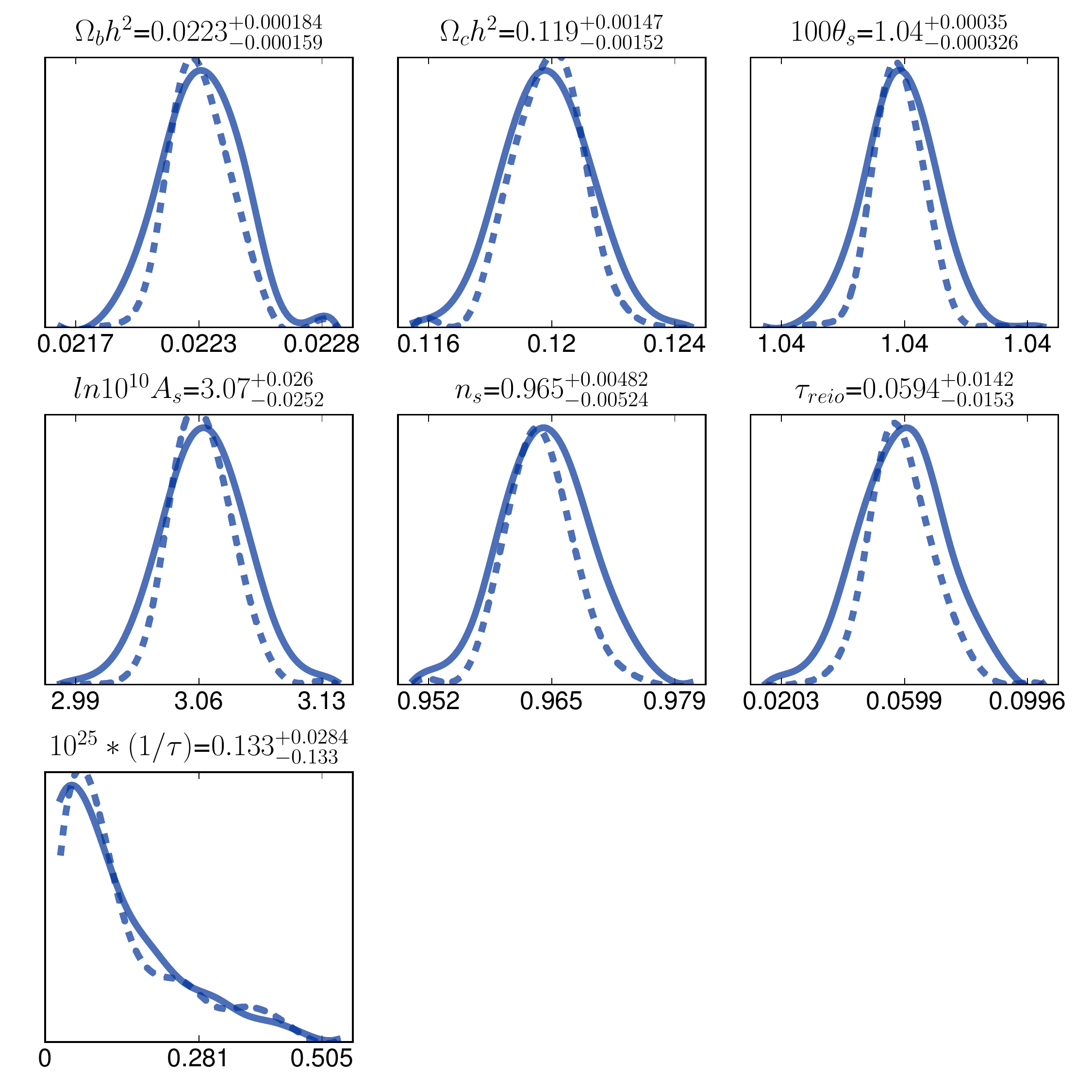} 
   \includegraphics[width=8cm]{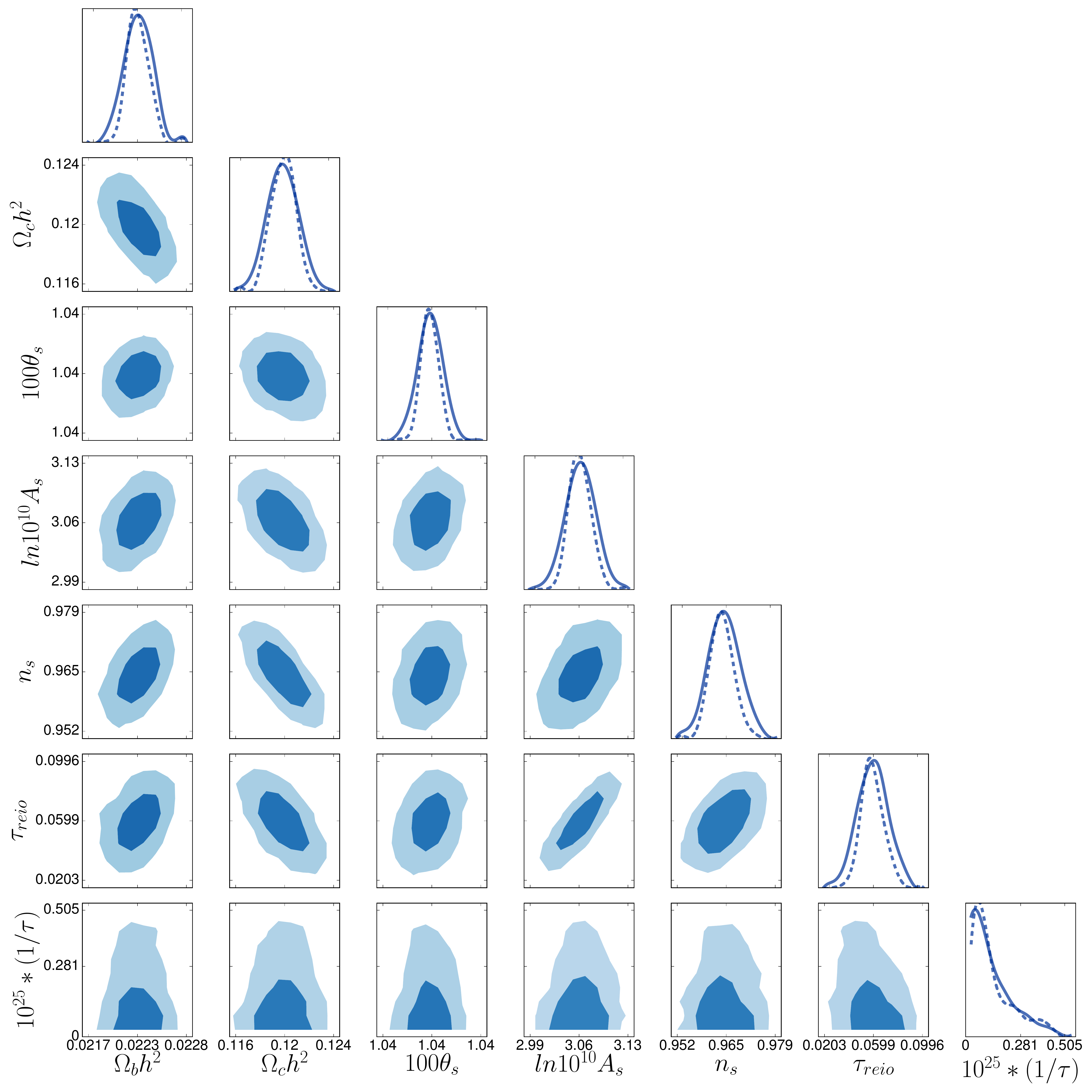}      
  \caption{
   The marginalized posterior probability distributions for the cosmological parameters  (upper panel), and the corresponding 2D joint probability distributions, including DM decay lifetime $\tau$ in units of s. For this example, we consider decay of DM with mass $2(m_e + 30) \text{MeV} \approx 60$ MeV, and assume the only decay channel is to $e^+ e^-$.}
\label{mcmc0}
\end{figure}

\begin{figure}
    \includegraphics[width=6.8cm]{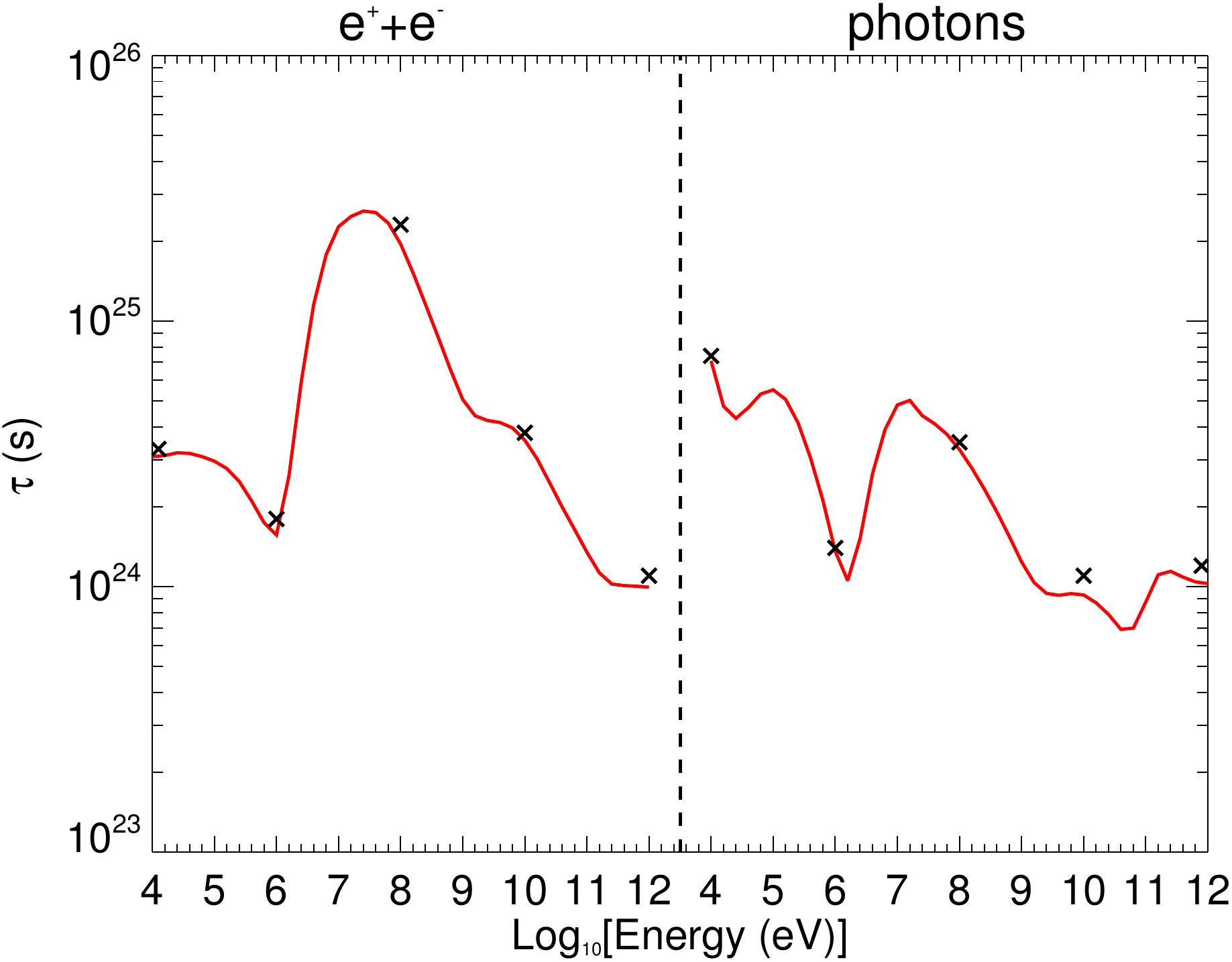}  
  \caption{Constraints on DM decay lifetimes from two methods: the MCMC constraint on decay to 30 MeV electrons and positrons, extrapolated to other energies using the first principal component (red line) and direct MCMC constraints (black crosses).}
\label{pc1mcmc}
\end{figure}

	In Table \ref{tab:mcmct}, we give the 95$\%$ C.L  lower limit on the DM decay lifetime, for different injection energies and species. In Fig. \ref{mcmc0}, we show the 1-D and 2-D posterior probability distributions for the cosmological parameters, in the case where we inject $e^+ e^-$ pairs 30 MeV of kinetic energy per particle. Comparing Tables \ref{tab:pcforecast} and \ref{tab:mcmct}, we find they are in good agreement with each other. These results are shown explicitly in Fig. \ref{pc1mcmc}. Typically, the true constraints are slightly weaker than the PCA-based forecasts; this is expected, as non-Gaussianity of the likelihood generally reduces significance / weakens constraints \cite{Verde:2009tu}, and any non-linearity will also tend to reduce the signal at larger energy injections. 
	
	We thus have confirmed that the first PC can be used to estimate correct limits on DM decay process. Furthermore, by calibrating the constraints to those from the MCMC and using the first principal component to translate the MCMC results to arbitrary models, we can cancel out most of the difference between the PCA and MCMC analyses, as we will discuss in the next section. 

\begin{table}
\def\arraystretch{1.5}
\begin{tabular}{c|P{2.0cm}|c}
 \hline
 species & energies  & decay lifetime / $10^{25}$ s (95 $\%$ CL)  \\  
 \hline 
 \multirow{5}{*}{electron} &
 10keV  & $0.33$   \\
 & 1MeV &   $0.18$  \\
 & 100MeV  & $2.31$ \\
 & 10GeV  &   $0.38$  \\
 & 1TeV   &$0.11$  \\ \hline
 \multirow{5}{*}{photon} &
  10keV   & $0.74$  \\
 & 1MeV & $0.14$ \\
 & 100MeV  & $0.35$ \\
 & 10GeV     &  $0.11$\\
 & 1TeV  &  $0.12$\\
 \hline
\end{tabular}
\caption{\emph{Planck} lower bounds on decay lifetime at 95 $\%$ confidence, using a MCMC analysis of actual data, for decays to $e^+ e^-$ pairs and photons at a range of energies. Here ``electron'' always labels the electron in an $e^+ e^-$ pair.}
\label{tab:mcmct}
\end{table}

\section{General constraints on DM decay}
\label{sec:discussion}

As we have shown in Eq. \ref{eq:coeffs}, any decaying DM model can be decomposed into a linear combination of the basis models with a set of coefficients $\{\alpha_i\}$, which in turn can be read off directly from its decay lifetime and the spectra of photons/electrons produced by its annihilation. The detectability of any DM model using the CMB anisotropy spectrum can be estimated by the dot product of this coefficient vector $\vec{\alpha}$ with the first PC; conversely, if no signal is seen, this dot product approximately controls the strength of the constraint on $p_\text{ref} = f_X/\tau$. Truncating Eq.~\ref{eq:constraint} to the first principal component, we can write the approximate forecast 95\% confidence limit as:  
\begin{align} & p_\text{ref} \lesssim \frac{2}{\sqrt{\lambda_1}} \frac{1}{\vec{\alpha} \cdot \vec{e}_1} = \frac{2}{\sqrt{\lambda_1}} \frac{p_\text{ref}}{\frac{f_X}{\tau} \vec{N} \cdot \vec{e}_1}, \nonumber \\
& \Rightarrow \tau \gtrsim f_X \frac{\sqrt{\lambda_1}}{2} \vec{N} \cdot \vec{e}_1, \end{align}
where $\vec{N}$ describes the spectrum of photons or electron/positron pairs (as appropriate to the basis model indexed by $i$) produced in a single decay,
\begin{equation} \vec{N} = \left\{ \frac{1}{M_\chi} E_i \frac{dN_{e^+ e^-,\gamma \gamma}}{d\ln E_i} \right\}, \quad i=1..N. \end{equation}
The sum over the elements of $\vec{N}$ should approximate the total fraction of the decaying-DM mass that proceeds into electrons, positrons and photons. (If only a small fraction of the DM mass decays into electromagnetically interacting channels, that is naturally captured in this formalism.)

Most of the discrepancy between the MCMC results and those of the PCA lies in the overall normalization, not in the shape of the first PC. Thus we can improve the PCA-based forecast by performing a single MCMC analysis for a reference model, and then using the PCA to predict the relative strength of constraints on other models. We will choose our reference model to correspond to injection of 30 MeV kinetic-energy electrons and positrons, i.e. $(\vec{N})_i = \delta_{ij}$ where $j$ indexes the basis model corresponding to injection of 30 MeV electrons and positrons, with a lifetime much longer than the age of the universe. Then if the MCMC constraint on this model for $f_X=1$ is $\tau < \tau_0$, for a general model we can estimate:
\begin{equation}\tau \gtrsim f_X \frac{\vec{N} \cdot \vec{e}_1}{\vec{e}_1(30 \, \text{MeV} \, e^+ e^-)} \tau_0. \end{equation}
In other words, we can use the first principal component plotted in Fig.~\ref{PCAfirst} to rescale constraints on the DM decay lifetime obtained from a MCMC analysis of a single reference DM model (chosen here to be long-lifetime DM decaying purely to $e^+ e^-$ pairs with kinetic energy per positron of 30 MeV). Note that the normalization of this principal component cancels out, only its shape is important. In analogy to the $f_\text{eff}$ parameter defined for annihilating DM in \cite{Slatyer:2015jla}, our ``detectability parameter'' $g_\text{eff}$ for a given model becomes:
\begin{equation} g_\text{eff} = \frac{\vec{N} \cdot \vec{e}_1}{\vec{e}_1(30 \, \text{MeV} \, e^+ e^-)}. \end{equation}
This parameter is proportional to $f_\text{eff}$, but has a different normalization due to the different reference model (the reference model for $f_\text{eff}$ was determined by the likelihood analysis already performed by the \emph{Planck} collaboration, and corresponded to 100\% power deposited into electrons/positrons/photons, with a redshift-dependent but energy-independent fraction of that power being promptly absorbed as hydrogen ionization). It is determined by the integral (or discrete sum) of the electron and photon spectra weighted by the first principal component.

For the \emph{Planck} data, we obtain the MCMC constraint on our reference model (decay to 30 MeV electrons and positrons) $\tau > \tau_0 = 2.6 \times 10^{25}$ s at 95\% confidence. (The corresponding PCA forecast limit is $3.25 \times 10^{25}$ s, using only the first PC.) Thus for general models we write:
\begin{equation} \tau \gtrsim f_X g_\text{eff} \times 2.6 \times 10^{25} \text{s} \quad \text{(95\% confidence)}. \end{equation}

To validate this approach, in Fig.~\ref{pc1mcmc} we compare two constraints on the DM lifetime for the models presented in Tables \ref{tab:pcforecast} and \ref{tab:mcmct}: (1) the directly computed MCMC bounds (Table \ref{tab:mcmct}), and (2) the MCMC bound on our reference model, extrapolated to other energies using the first PC (this is equivalent to rescaling all the results in Table \ref{tab:mcmct} by a constant, determined by the comparison between the MCMC result and PCA forecast for the reference model). In this case, since we are assuming $f_X=1$ for all models and considering models which produce only $e^+ e^-$ pairs or photons at a specific energy, the bound on the lifetime is directly proportional to the first principal component (Fig.~\ref{PCAfirst}). We find good agreement, at the $\sim 10\%$ level, for all points tested. 

We then apply this approach to DM decay to SM particles, considering 28 decay modes for DM masses from 10 GeV to 10 TeV; the resulting spectra of photons and $e^+ e^-$ pairs are provided in the PPPC4DMID package \cite{Cirelli:2010xx}. We assume that 100\% of the DM is decaying, with lifetime much longer than the age of the universe. 

 We also provide constraints on DM below 10 GeV decaying to photons and $e^+ e^-$ pairs, the latter either as a direct decay, or via decay to a pair of unstable mediators (denoted $VV$) which each subsequently decay to an $e^+ e^-$ pair.
 
\begin{figure}
  \includegraphics[width=8cm]{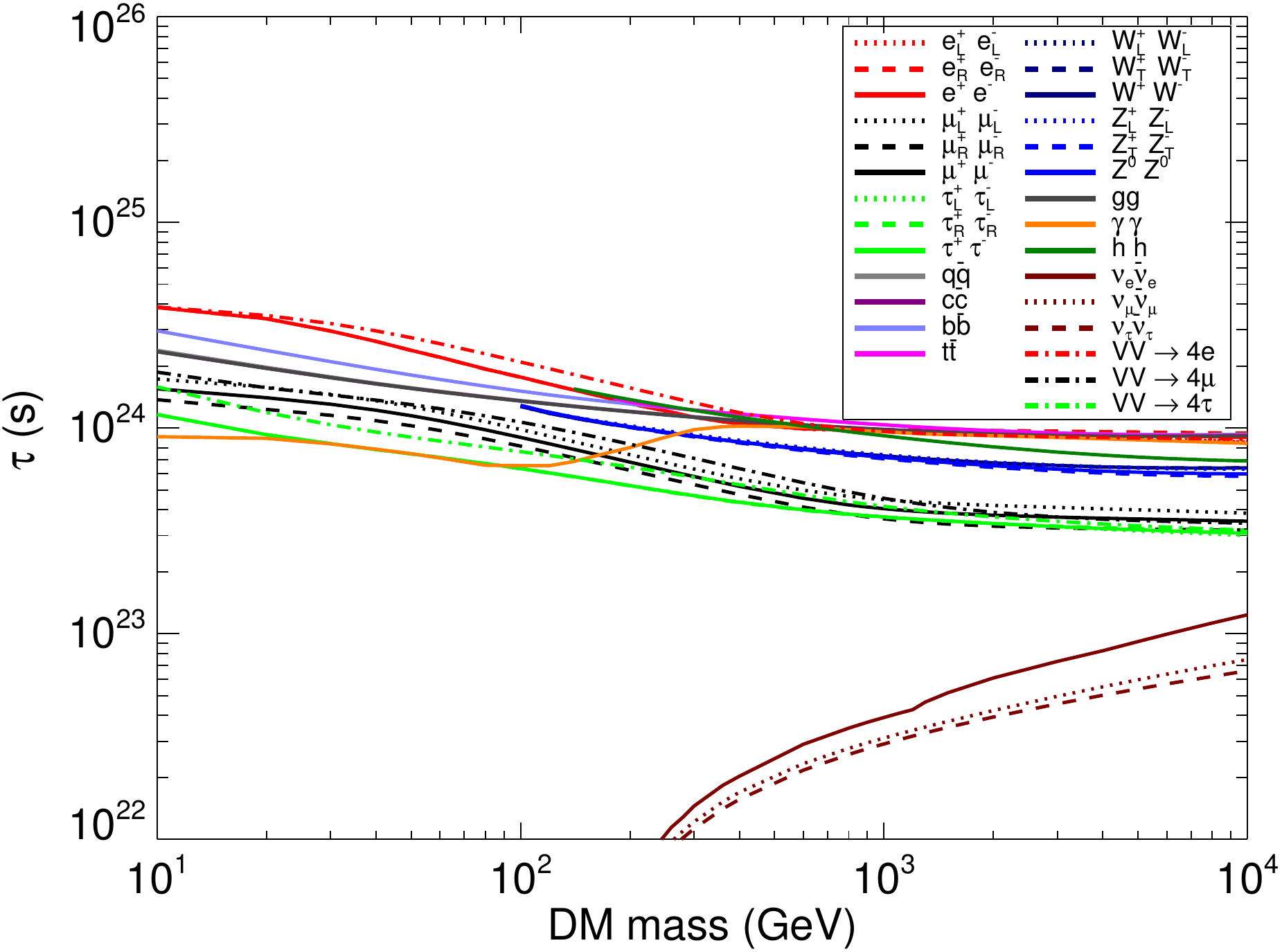}\\[3ex]
  \includegraphics[width=8cm]{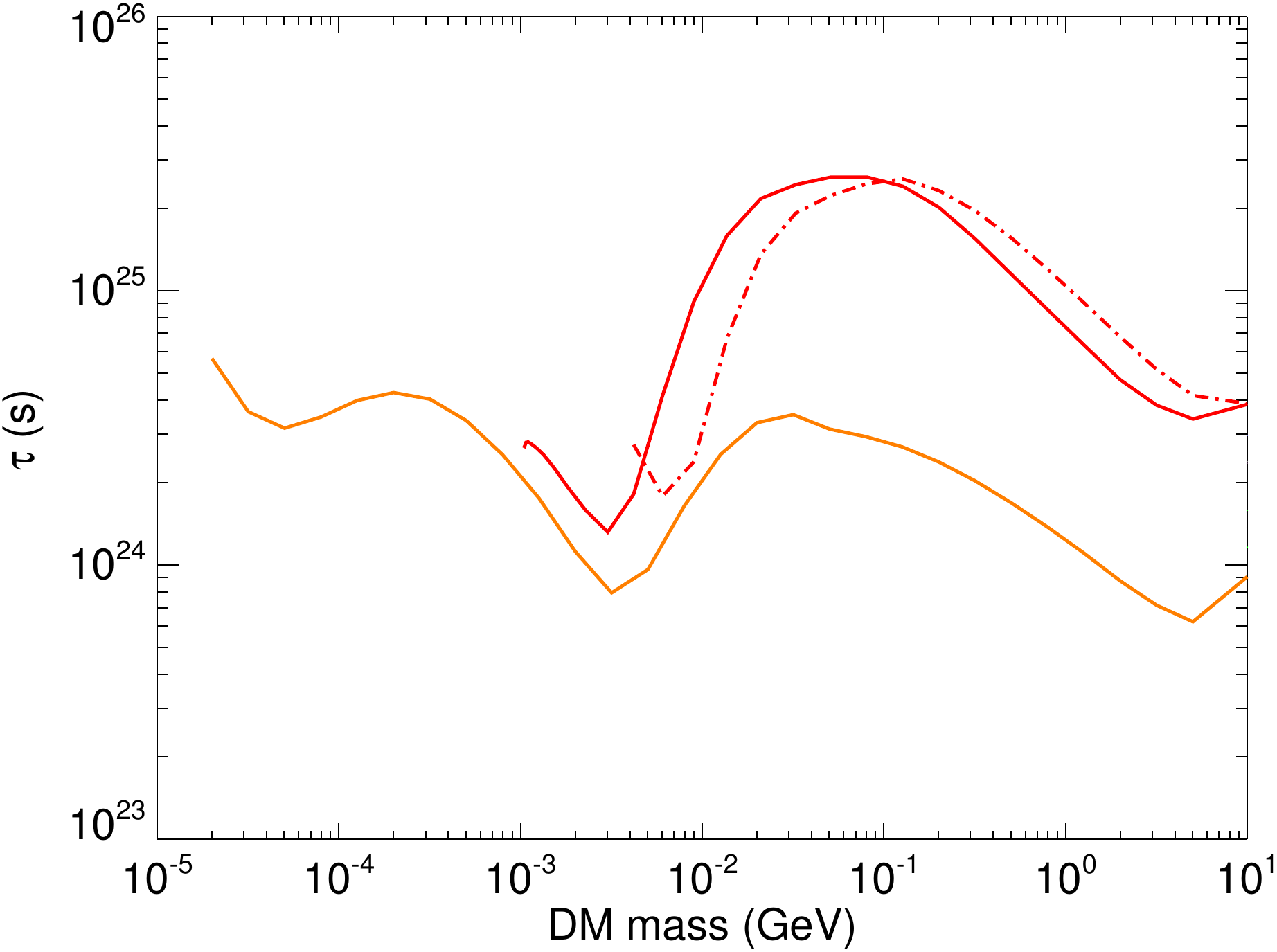}
\caption{The estimated lifetime constraints on decaying DM particles, from PCA for \emph{Planck} calibrated to the MCMC result for our reference model (injection of 30 MeV electrons/positrons). The upper panel covers the DM mass range from 10 GeV to 10 TeV. The lower panel covers the range from keV-scale DM masses up to 10 GeV for the $e^+e^-$, $\gamma\gamma$ and VV $\rightarrow$ 4$e$ channels.}
\label{dmconstraint}
\end{figure}

 The resulting constraints on the lifetime are shown in Fig. \ref{dmconstraint}. We note several salient points:
\begin{itemize}
\item The label q = u,d,s denotes a light quark and h is the SM Higgs boson. The distinction between polarization of the leptons (Left- or Right-handed fermion) and of the massive vectors (Transverse or Longitudinal) matter for the electroweak corrections. The last three channels denote models in which the DM decays into a pair of intermediate vector bosons VV, which then each decay into a pair of leptons.
\item Decays to neutrinos are the least constrained, and are only constrained at all at high masses, as the only photons and $e^+ e^-$ pairs in these decays are produced through electroweak corrections (e.g. final state radiation of electroweak gauge bosons). These limits are $\sim 2-3$ orders of magnitude weaker than present-day indirect searches using neutrino telescopes \cite{Esmaili:2012us}.
\item Other SM final states populate a band of decay-lifetime constraints whose vertical width is roughly a factor of 4-5.
\item In contrast to annihilating DM, the detectability function is quite sharply peaked around $\sim 100$ MeV electrons/positrons, for decaying DM. Consequently, channels that produce copious soft electrons/positrons can have enhanced detectability -- this is in contrast to the usual situation for indirect searches in the present day, where softer spectra are typically more difficult to detect due to larger backgrounds.
\item For TeV DM and above, the contributions from the electron/positron and photon spectra are typically comparable, and the detectability depends primarily on the total power proceeding into electromagnetic channels.
\end{itemize}

One might ask how these constraints compare to existing bounds. For long-lifetime decaying DM, there are stringent constraints on the decay lifetime from a wide range of indirect searches (e.g. \cite{Yuksel:2007dr,PalomaresRuiz:2007ry,Zhang:2009ut,Cirelli:2009dv, Bell:2010fk,Dugger:2010ys,Cirelli:2012ut,1475-7516-2012-10-043,Essig:2013goa,Mambrini2016807}). In general, these constraints are considerably stronger than our limits, probing lifetimes as long as $10^{27-28}$ s. The exception is for MeV $-$ GeV DM decaying to $e^+ e^-$ pairs; these pairs are difficult to detect directly. They do produce photons via internal bremsstrahlung and final state radiation, and in \cite{Essig:2013goa}, data from HEAO-1, INTEGRAL, COMPTEL, EGRET, and the \textit{Fermi Gamma-Ray Space Telescope} (\textit{Fermi}) were used to set constraints on such decays by searching for these photons. These constraints are conservative in that they subtract no astrophysical background model, but they do assume a Navarro-Frenk-White \cite{Navarro:1995iw} density profile for the DM.

\begin{figure}
    \includegraphics[width=7cm]{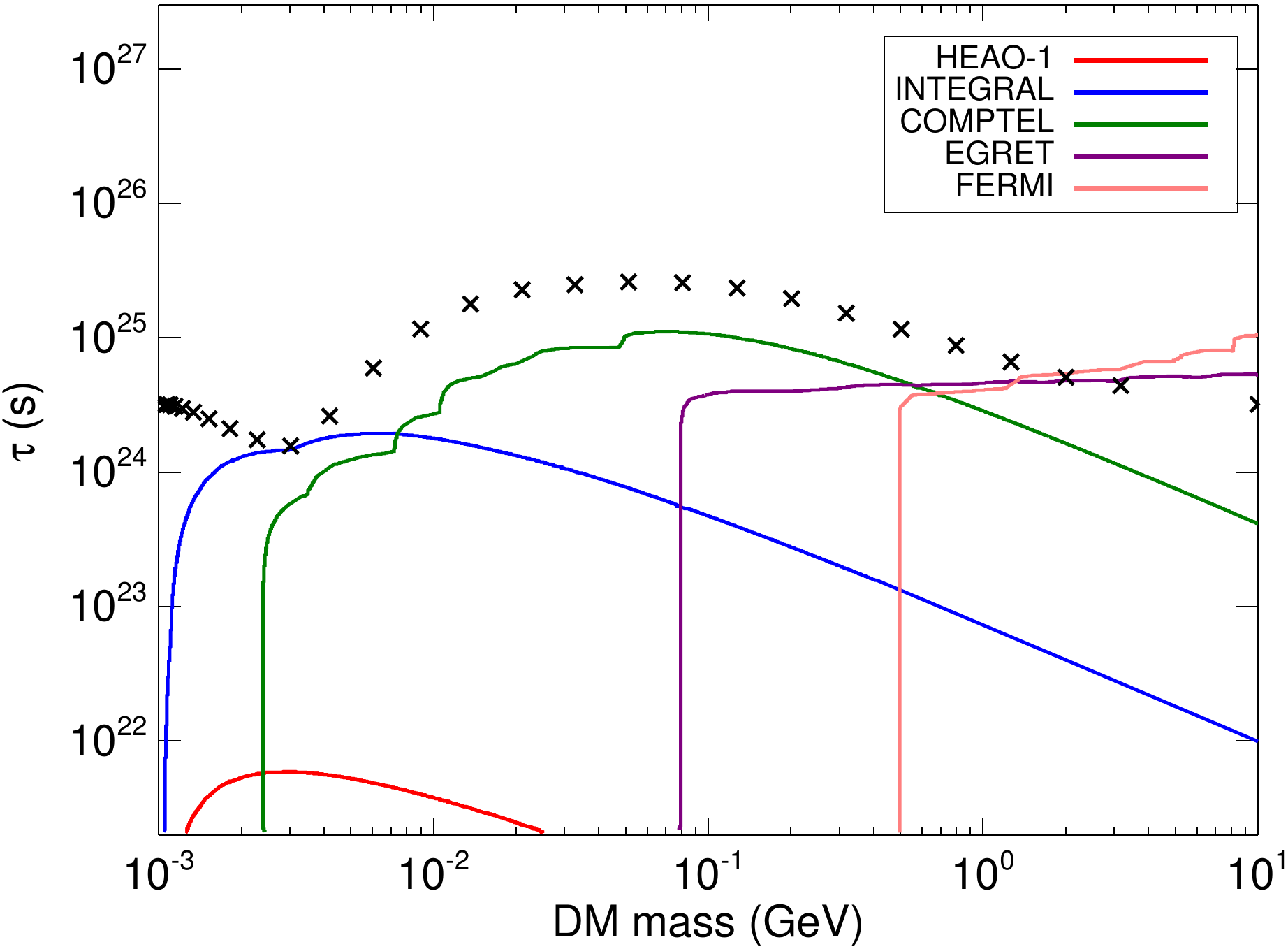}   
  \caption{
   Lower bounds on the DM decay lifetime, for decay to $e^+e^-$, from present-day diffuse photon searches (colored lines) and from our results by using PCA (black crosses) calibrated to the MCMC bound for our reference model.}
\label{compare}
\end{figure}

In Fig.~\ref{compare} we compare our CMB constraints (which are of course independent of any assumptions about the halo DM density) to these limits. In the MeV $-$ GeV mass range, our limits exceed the previous best bounds on the decay lifetime by a factor of a few. 

One might ask how much these bounds have the potential to improve. As shown in Fig.~\ref{PCAfirst}, the shape of the first PC is very similar for WMAP7, \emph{Planck} and an experiment that is CVL up to $\ell=2500$. Thus the main effect of moving closer to a CVL experiment would be to improve the constraints on all channels by a constant factor. Examining the eigenvalues of the first PC in the \emph{Planck} and CVL cases, we expect the limit to improve by a factor of $\sim 5$ with an experiment that is CVL up to $\ell=2500$.

Let us now discuss the case where a small mass fraction of the DM decays \emph{prior} to the present day. This immediately removes most limits from present-day indirect searches. Limits from structure formation, in the case where the decay is from a metastable excited state of DM and thus confers a velocity kick on the remaining DM, can constrain decays with lifetimes $\sim 10^{16}$ s (e.g. \cite{Peter:2010au, Peter:2010jy, Peter:2010sz, Peter:2010xe}). At lifetimes much shorter than $\sim 10^{12-13}$ s, limits from Big Bang Nucleosynthesis will generally dominate (for one example scenario, see \cite{Cyburt:2009pg}). However, in the lifetime range $\sim 10^{13-16}$ s, limits from the CMB are uniquely powerful \cite{Slatyer:2012yq}.

\begin{figure}
    \includegraphics[width=6.5cm]{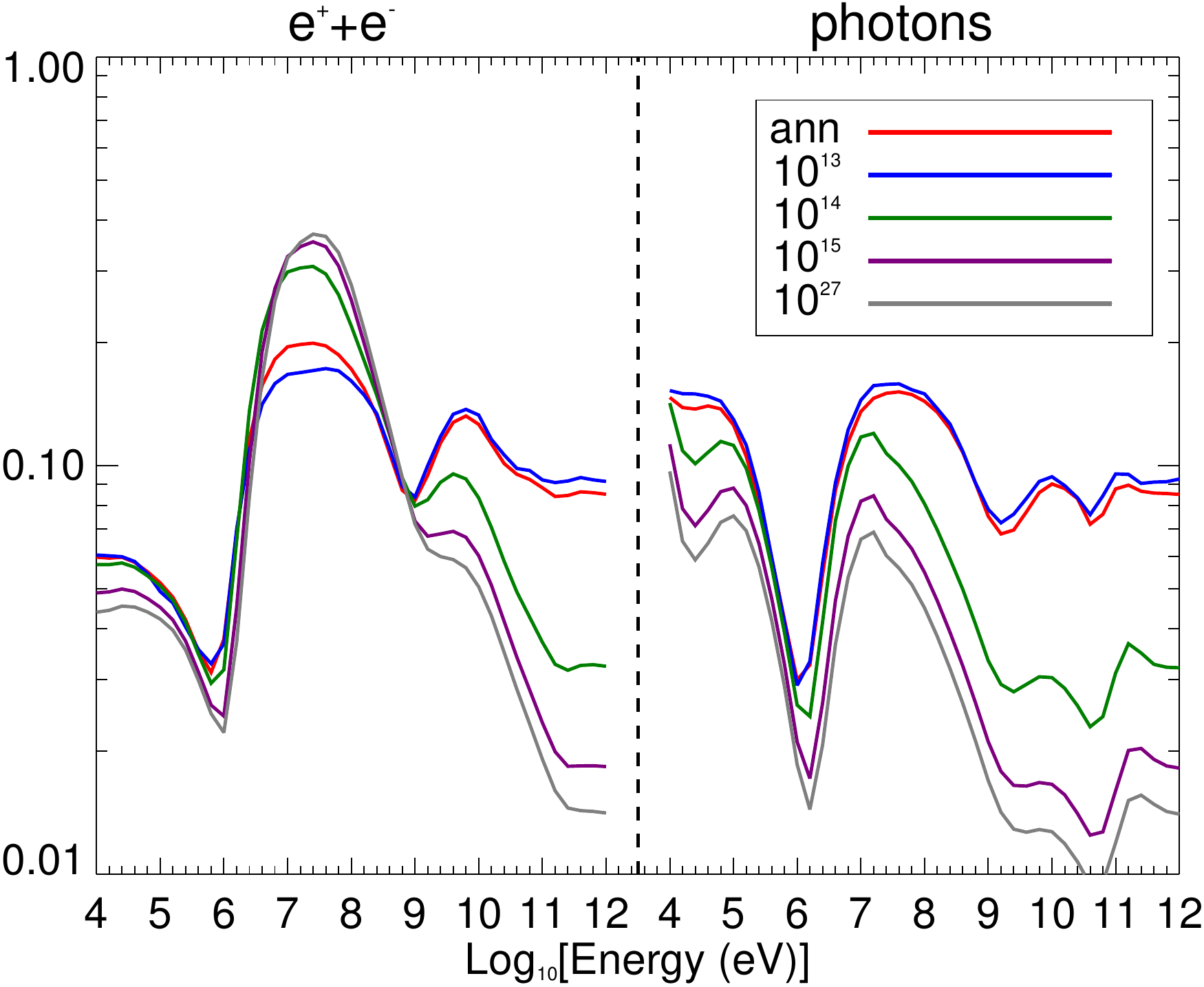}  
  \caption{
   First principal components for \emph{Planck} for annihilation (red), decays with fixed short lifetimes of $10^{13}$ s (blue), $10^{14}$ s (green) or $10^{15}$ s (purple), and long-lifetime ($10^{27}$ s) decays (gray).}
\label{fig:shortlifetimepca}
\end{figure}

In this case, our PCA must be extended to account for shorter lifetimes. Fig.~\ref{lifetime} allows us to approximately translate the MCMC limits on long-lifetime DM, decaying to 30 MeV electrons/positrons, into limits on the same decay channel but for shorter lifetimes. We can then perform PCA holding the lifetime fixed but varying the energy of injection and injected species, as in the case of long-lifetime DM, to translate these bounds into limits on other channels at the same lifetime. We show the resulting $\vec{e}_1$ curves in Fig.~\ref{fig:shortlifetimepca}. The analogous curves can be obtained for intermediate lifetimes by interpolation. 

It is interesting to note that as the decay lifetime becomes shorter, the first PC comes to resemble that for annihilation; this is because the difference in the PCs between long-lifetime decay and annihilation arises from the different redshifts at which the main contribution to the signals occur. As the decay lifetime is shortened, more of the signal originates from higher redshifts, and the PC for decay becomes more similar to that for annihilation (a redshift of 600, where the contribution to the annihilation signal peaks, corresponds to a cosmic age of $\sim 3 \times 10^{13}$ s).

\begin{figure}
    \includegraphics[width=6.8cm]{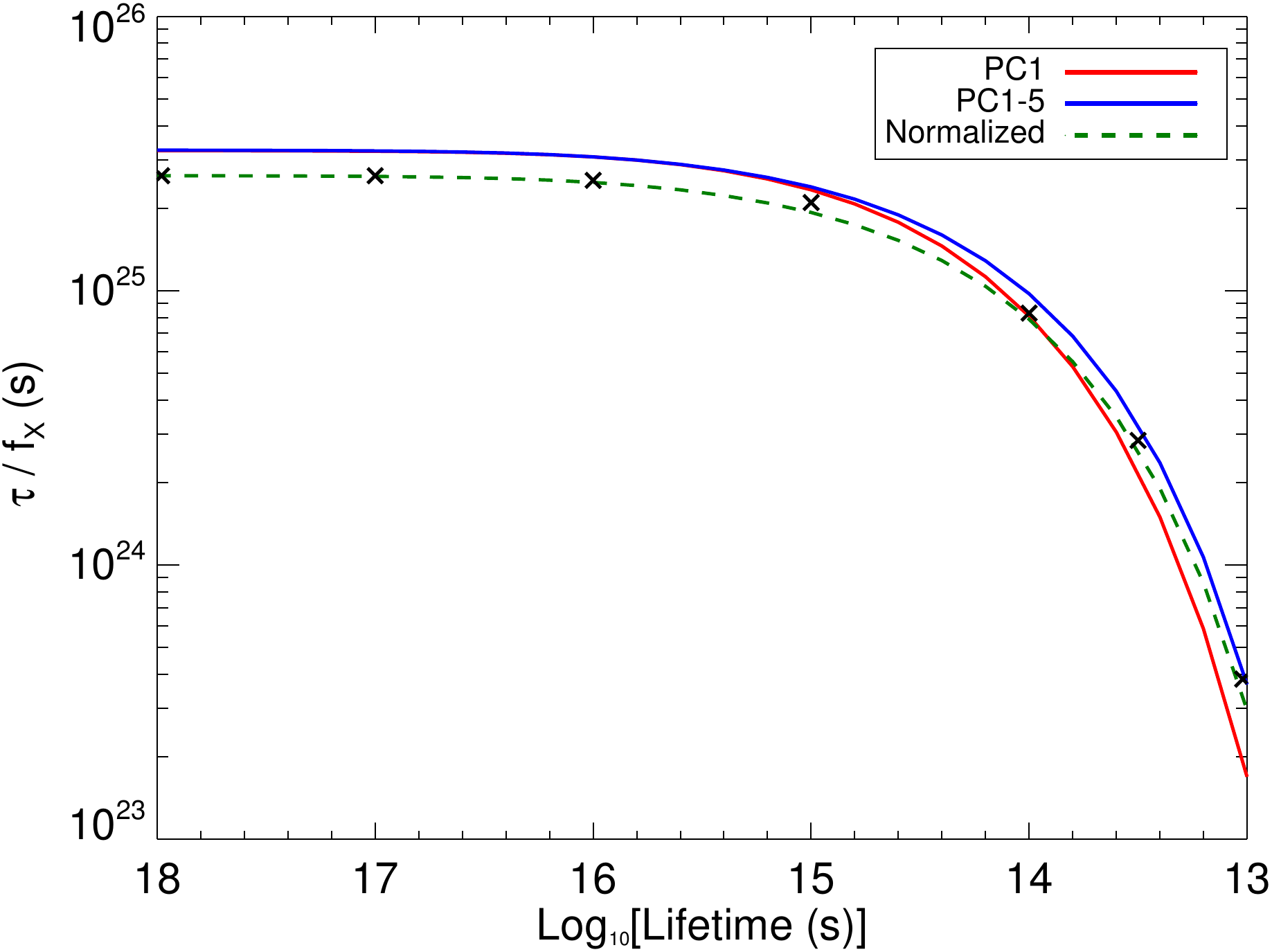}  
  \caption{The forecast bound on decaying DM properties, obtained by using different numbers of PCs. The red line uses the first PC and blue line uses the first 5 PCs; the contribution from higher PCs is negligible.  MCMC results are shown with black crosses. The green dashed line is the result of the blue line normalized to the MCMC result for a lifetime of $10^{18}$ s.}
\label{lifetimemcmc}
\end{figure}

\begin{figure}
    \includegraphics[width=7.2cm]{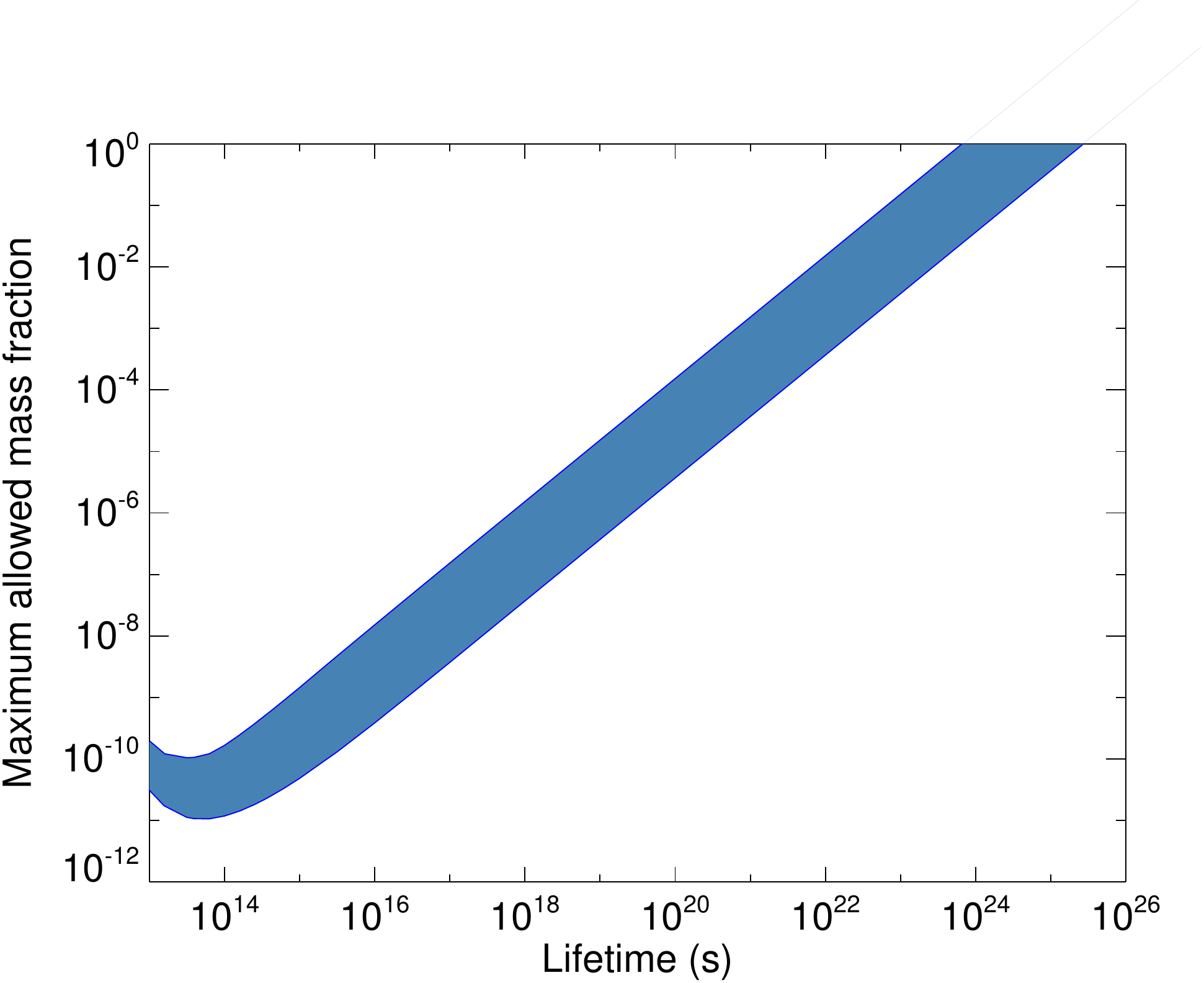}   
  \caption{
   Range of upper bounds on the mass fraction of DM that can decay with a lifetime $\tau$, for injections of 10 keV $-$ 10 TeV photons and $e^+ e^-$ pairs; the width of the band represents a scan over injection species and energy. The constraint is based on the PCA (first PC only) calibrated to the MCMC bound for our reference model.}
\label{shortconstraints}
\end{figure}

It is worthwhile to note that if the first PC is suppressed, the high PCs could give a sizable contribution to the forecast constraint. We show in Fig.~\ref{lifetimemcmc} the constraints obtained by using different numbers of PCs. For the short-lifetime DM, the correction from the high PCs becomes important. The MCMC results in this plot show that the difference between the PCA prediction and MCMC result is not a constant ratio with respect to lifetime. We therefore normalized the PCA result (summing up the higher PCs) to the MCMC result for $\tau = 10^{18}$ s, and used this normalized curve to estimate the constraints on short-lifetime decays. The resulting constraint is slightly weaker than we would obtain using the full MCMC for the shortest lifetimes we test; thus our constraints on short lifetimes will be conservative.

In Fig.~\ref{shortconstraints} we show the resulting estimated bounds on the mass fraction of DM that can decay, as a function of decay lifetime, based on the 2015 \emph{Planck} data. Rather than show results for individual models (which would require a scan over DM mass and annihilation channel), we simply show the band traced out by injection of $e^+ e^-$ pairs and photons at 10 keV $-$ 10 TeV energies. This figure updates Fig.~8 of \cite{Slatyer:2012yq}. Note that our limits weaken more rapidly than the bounds in \cite{Slatyer:2012yq} as the decay lifetime becomes shorter than the age of the universe at recombination (i.e. $\sim 10^{13}$ s); we attribute this to the fact that \cite{Slatyer:2012yq} used an older prescription for the fraction of power proceeding into ionization, which significantly overestimated the power into ionization when the background ionization level is non-negligible (as is the case during and shortly after recombination) \cite{Galli:2013dna}. 

\section{Conclusion}
\label{sec:conclusion}

Using principal component analysis, we have demonstrated that the imprint of general models of decaying DM on the CMB anisotropy spectrum -- via changes to the ionization and temperature history -- can be approximately described by a single parameter, if the lifetime of the DM is much longer than the age of the universe. After performing a detailed likelihood analysis on a single model to calibrate the constraints, which we have done using \emph{Planck} 2015 data, limits on the decay lifetime for all other models can be determined by a simple integral of the photon/electron spectra from annihilation products, weighted by the first principal component. Including higher principal components changes the decay lifetime constraints by less than $10\%$ in most cases, and we have validated our approach with MCMC studies.
	
	We find lifetime constraints typically of the order of $10^{25}$ s. These constraints outperform limits from the Galactic diffuse emission for MeV $-$ GeV DM annihilating primarily to $e^+ e^-$ pairs (or to particles which decay dominantly to $e^+ e^-$). More generally, they provide a robust limit on decay lifetime for a very wide range of models, evading any uncertainties associated with astrophysical backgrounds or the DM density distribution.
	
	We can also constrain the decay of a subdominant DM species, or a metastable state of DM, with lifetimes much \emph{shorter} than the current age of the universe, so long as the lifetime exceeds $\sim 10^{13}$ s (roughly the age of the universe at recombination). For shorter lifetimes, the constraints weaken drastically, and numerical issues limit our ability to compute even these weakened bounds; it is likely that for lifetimes much shorter than $10^{13}$ s, constraints from distortions of the CMB energy spectrum or modifications to Big Band Nucleosynthesis will be stronger than those computed with our current approach. 
	
	For such short lifetimes, only a tiny fraction of the total mass density of DM can decay, either because each decay liberates only a small fraction of the original particle's energy, or because the decaying species is only a small fraction of the total DM. We set upper limits on the mass fraction of DM that can decay as strong as $10^{-11}$, for lifetimes $\sim 10^{14}$ s.

\acknowledgments

We thank Hongwan Liu, Lina Necib, Nicholas Rodd, Ben Safdi, and Wei Xue for helpful discussions, and Bhaskar Dutta for a conversation that stimulated this project. This work is supported by the U.S. Department of Energy under grant Contract Numbers $\text{DE-SC00012567}$ and $\text{DE-SC0013999}$. Wu is partially supported by the Taiwan Top University Strategic Alliance (TUSA) Fellowship. 

\onecolumngrid

\newpage

\begin{appendix}
\section{Subsequent Principal Components}
\label{sec:threepcs}

In this appendix we display the second and third principal components, in addition to the first PC displayed in the main text. 

 \begin{figure}[h]
    \includegraphics[width=5.5cm]{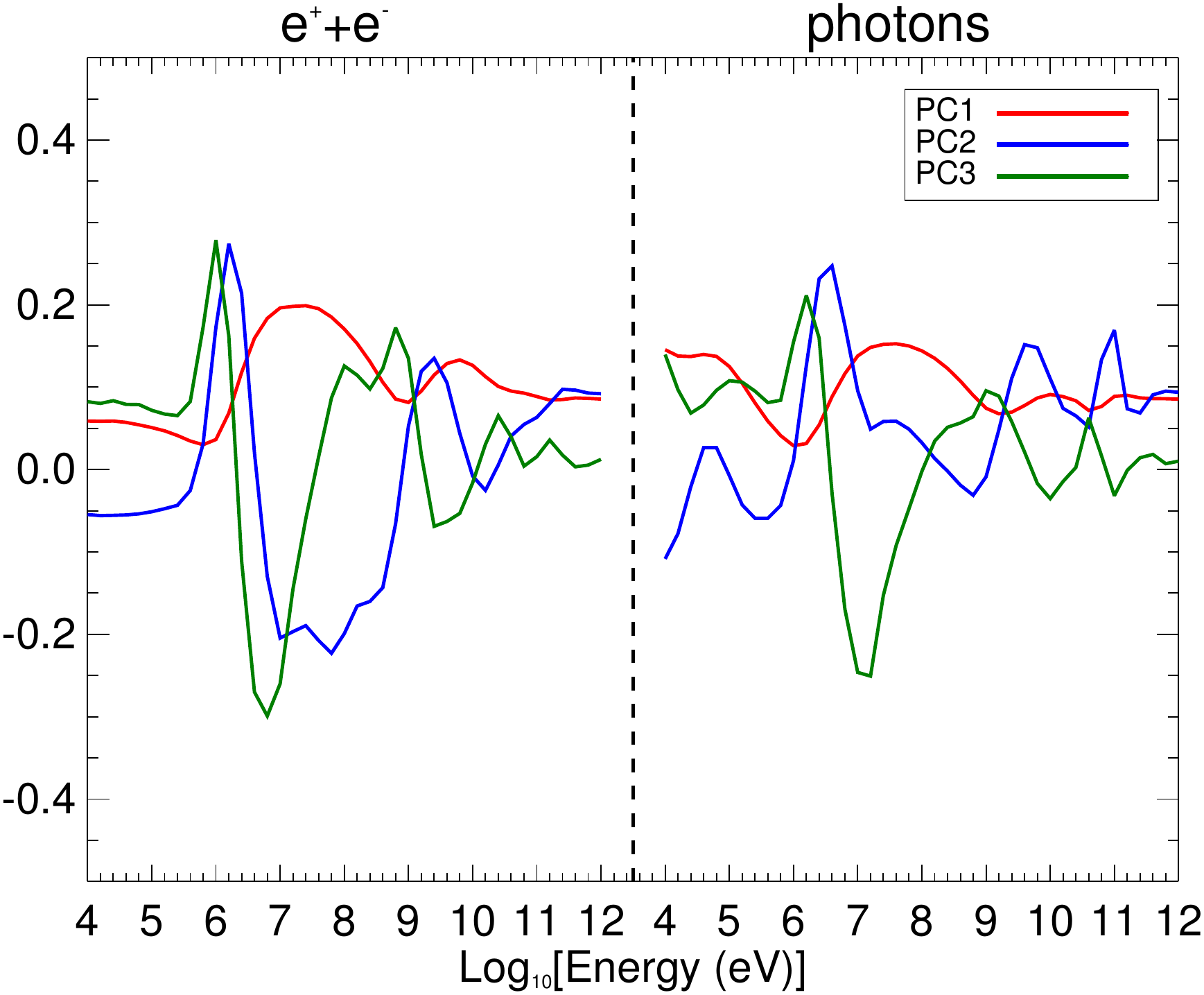}
   \includegraphics[width=5.5cm]{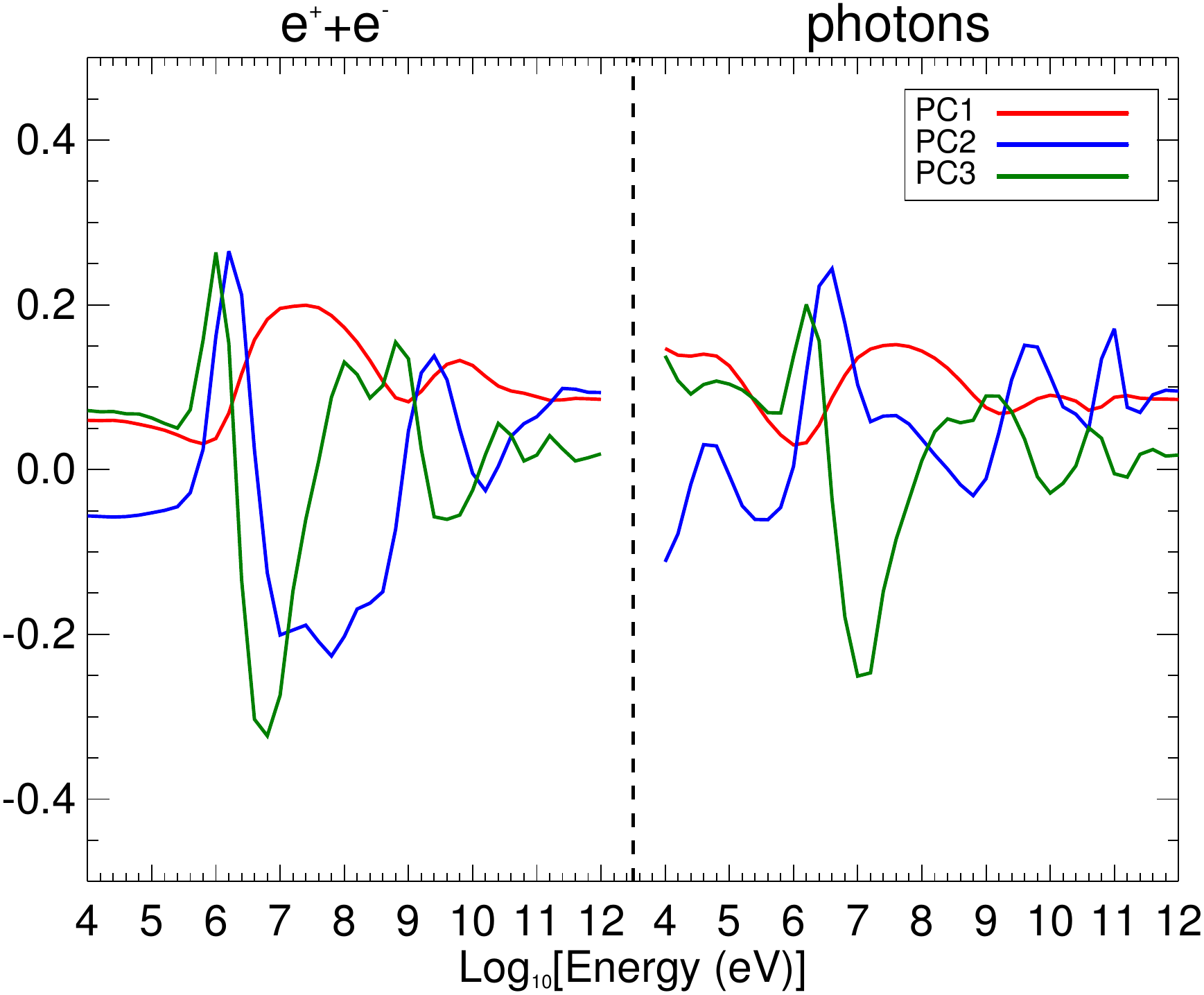} 
   \includegraphics[width=5.5cm]{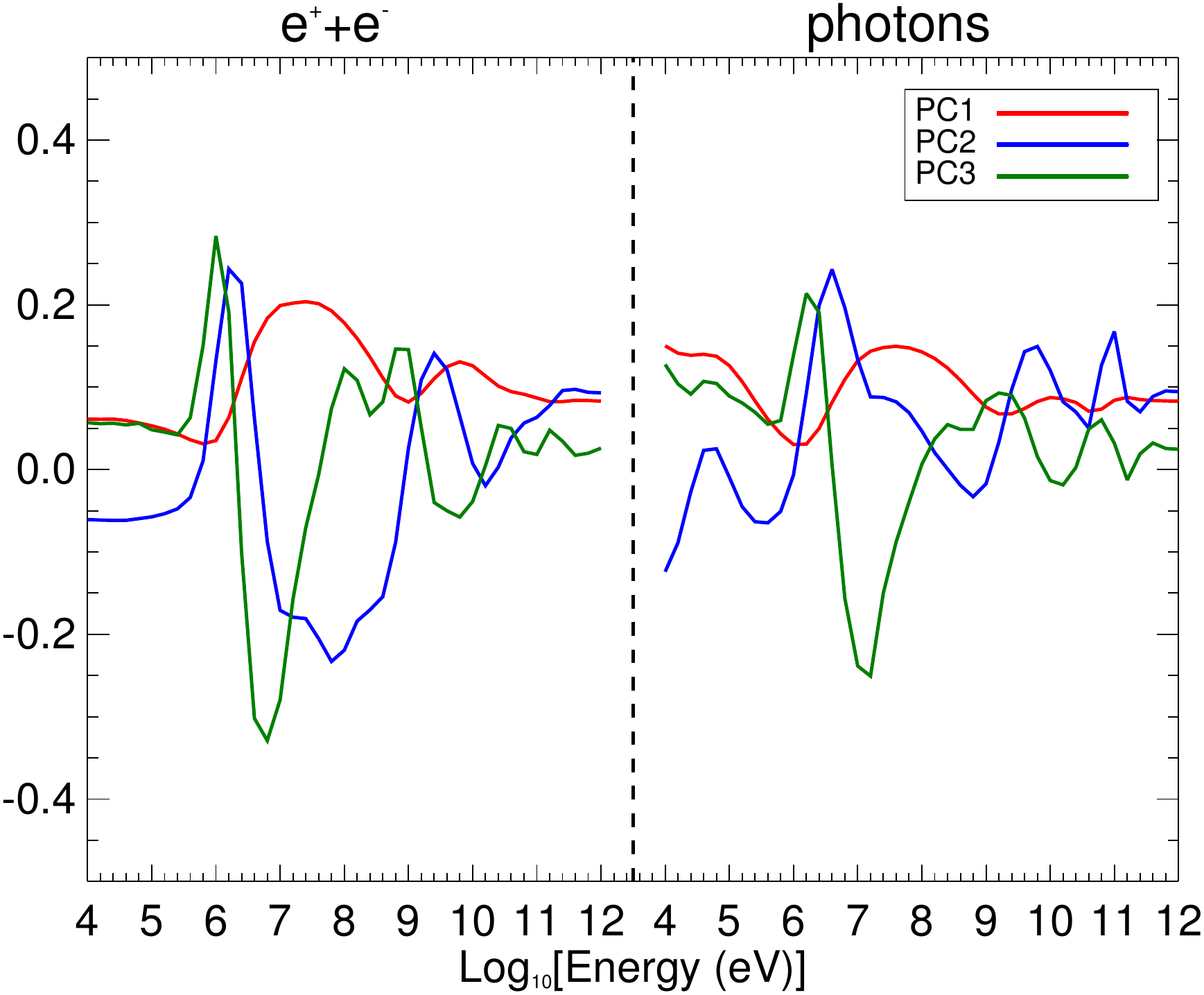} \\
   \includegraphics[width=5.5cm]{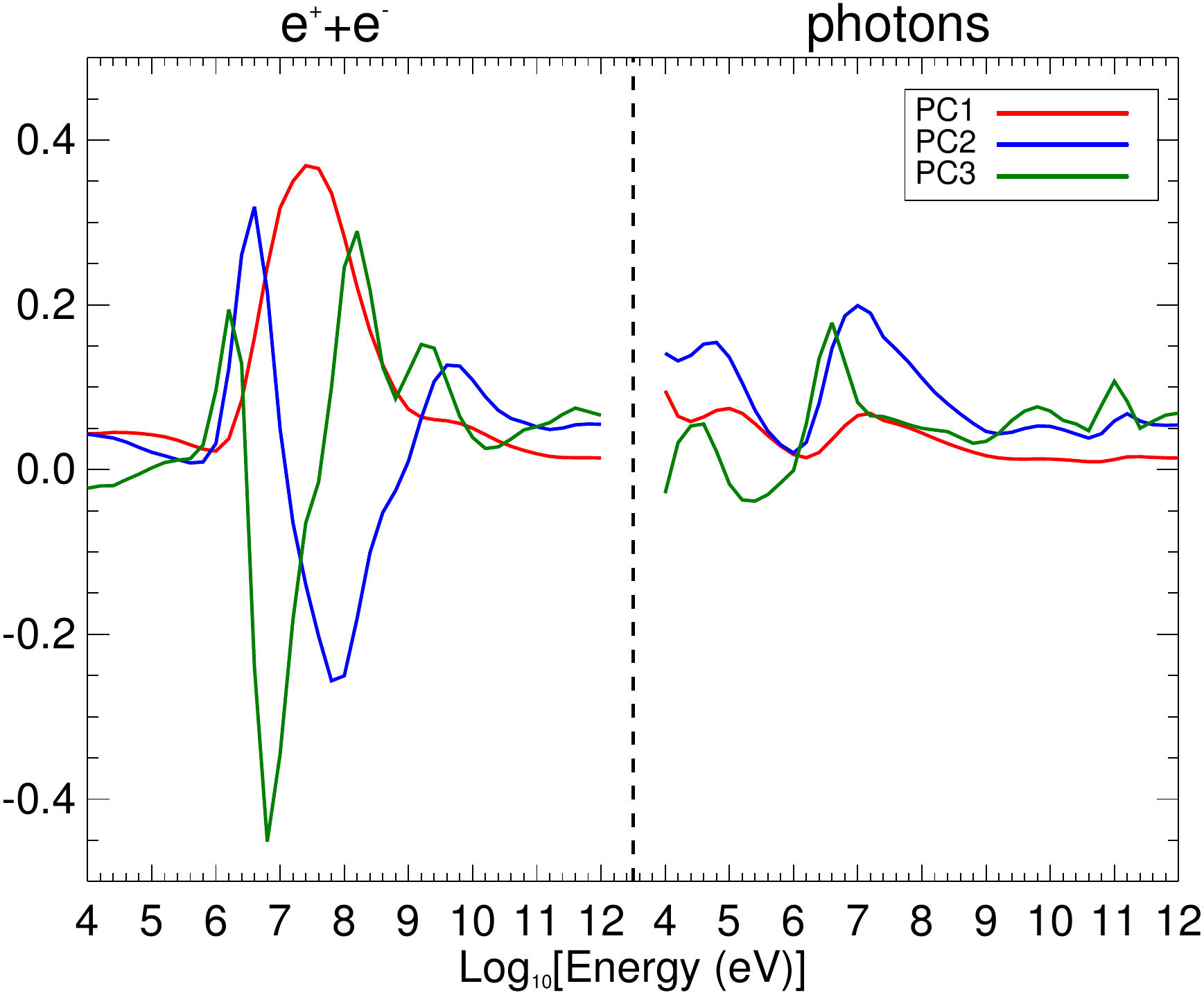}
   \includegraphics[width=5.5cm]{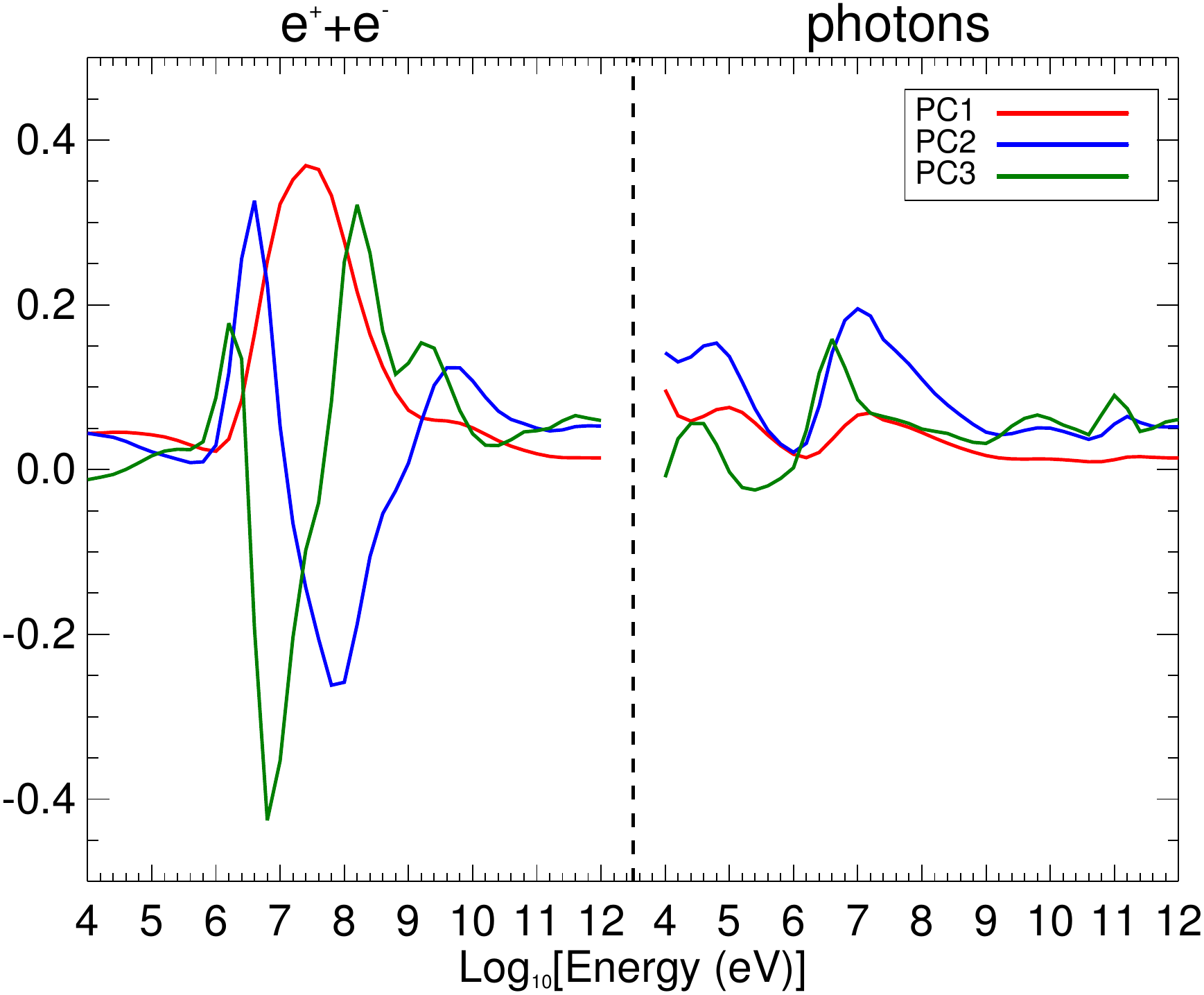} 
   \includegraphics[width=5.5cm]{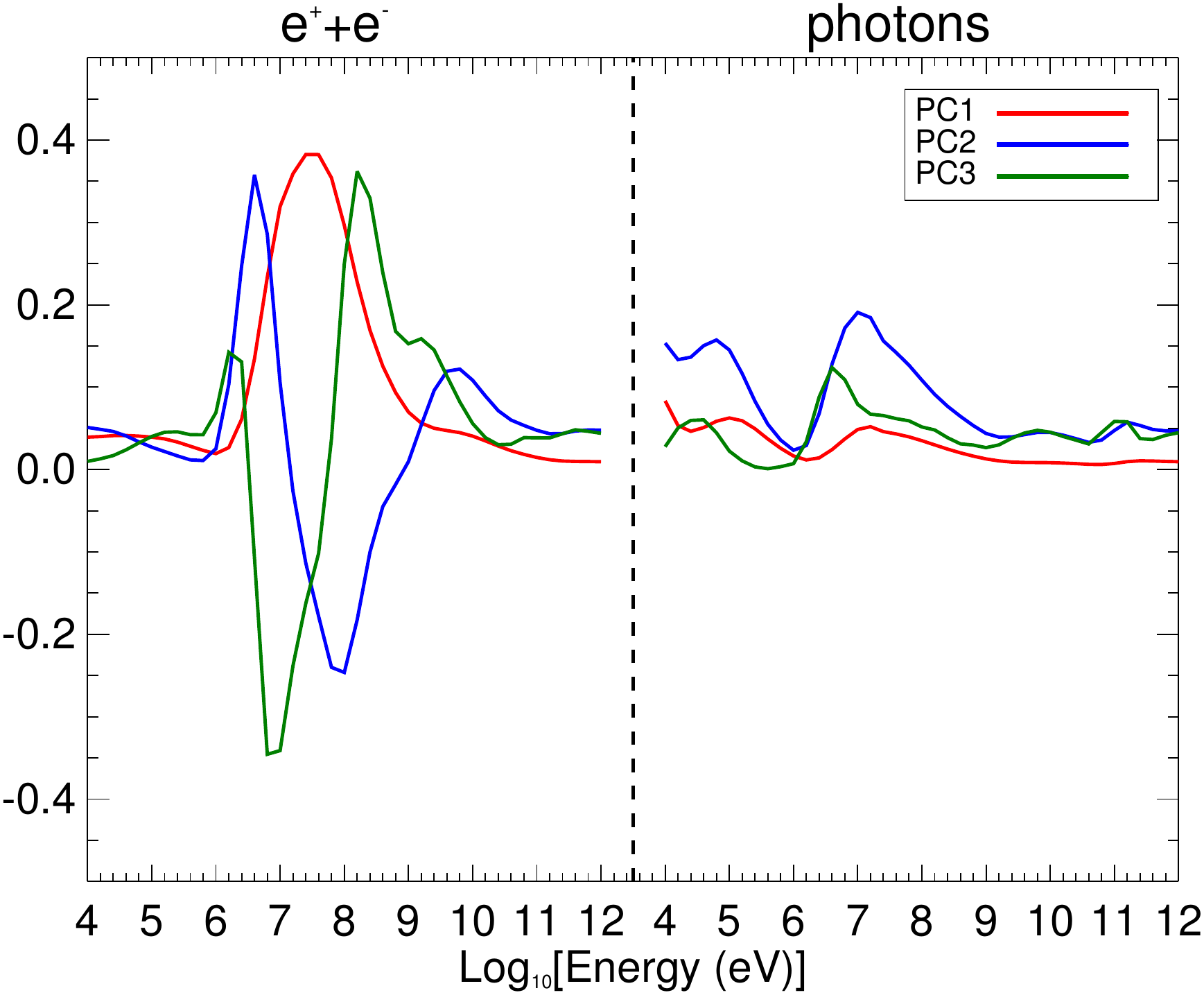} \\
   \includegraphics[width=5.5cm]{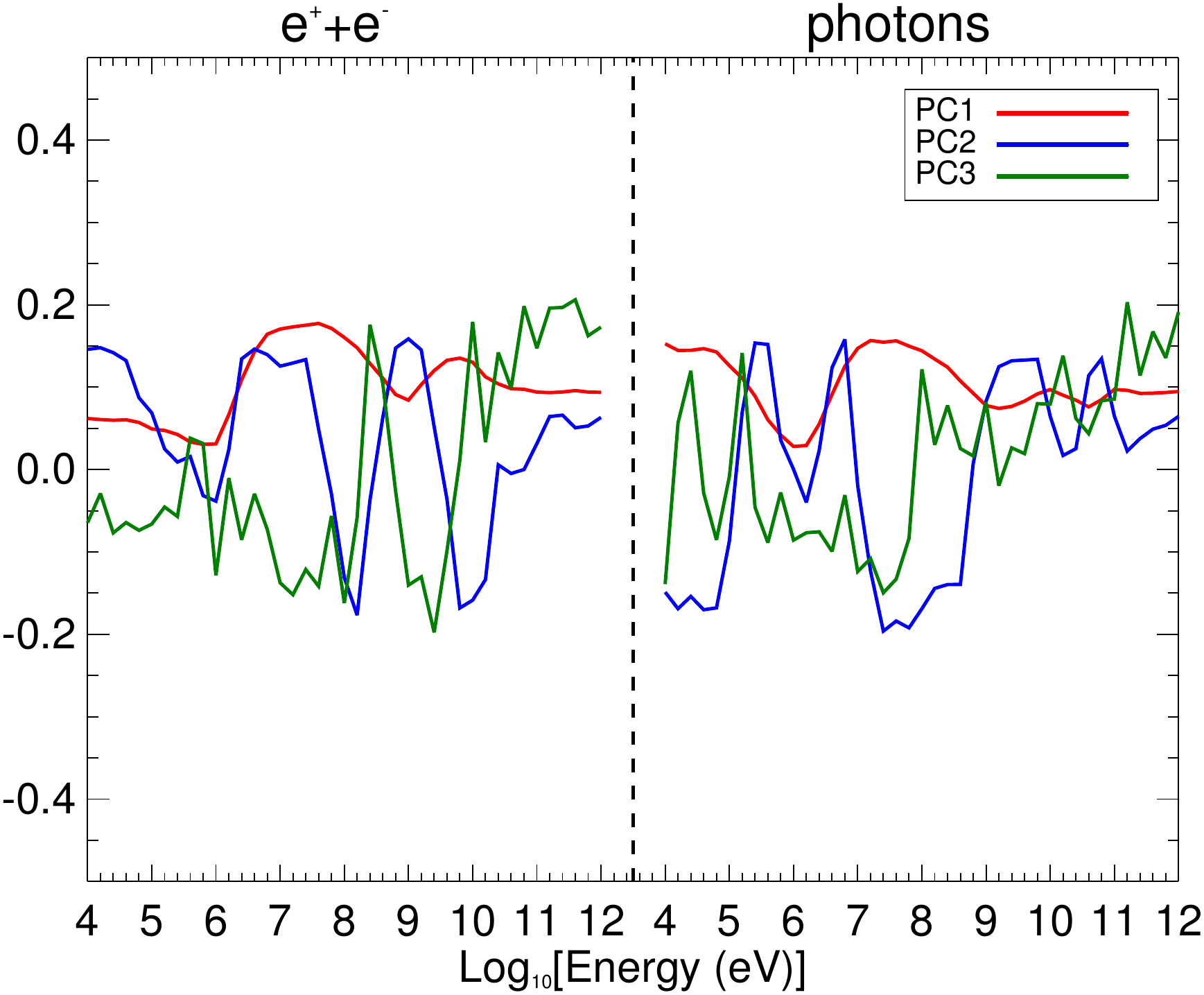}
   \includegraphics[width=5.5cm]{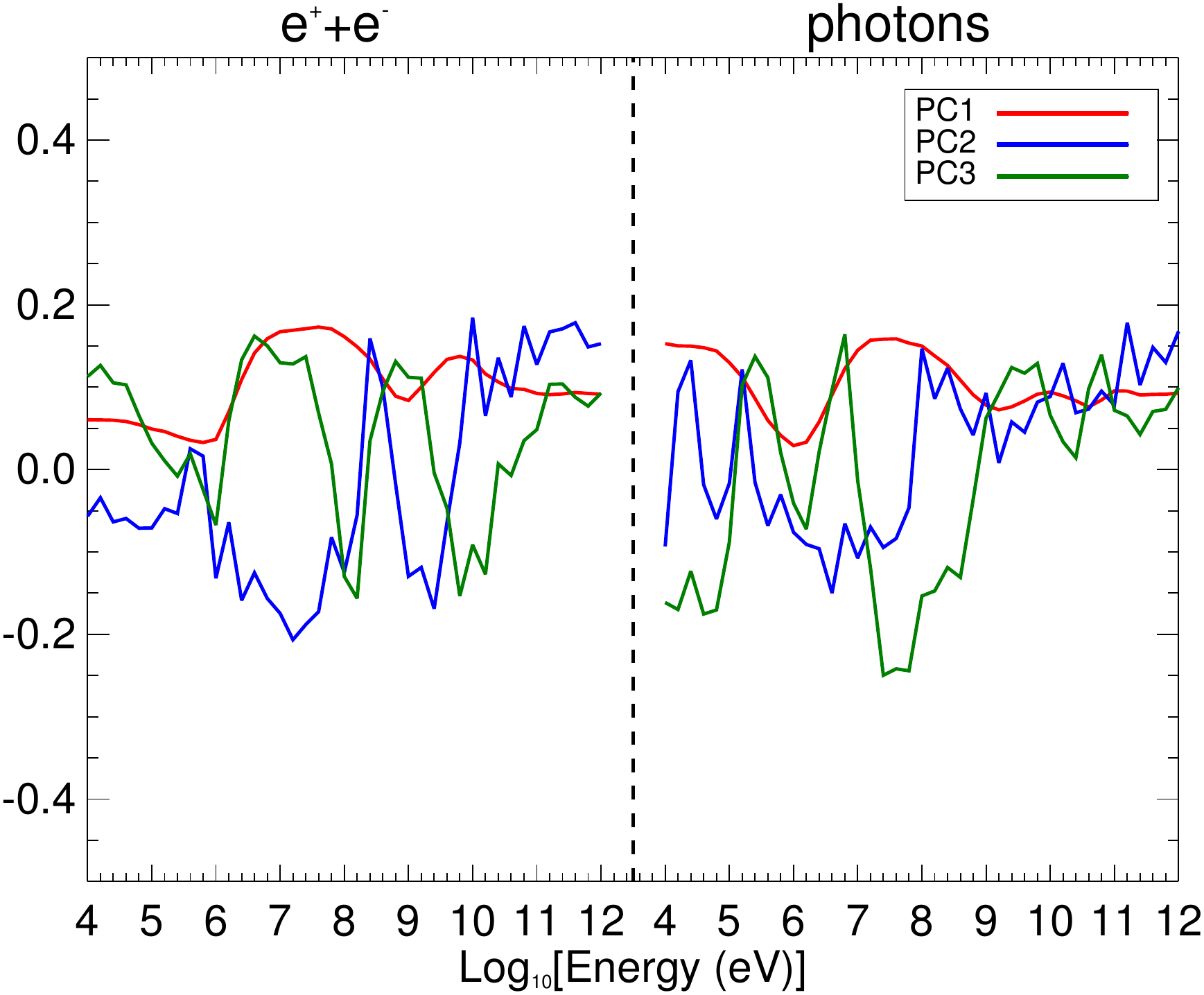} 
   \includegraphics[width=5.5cm]{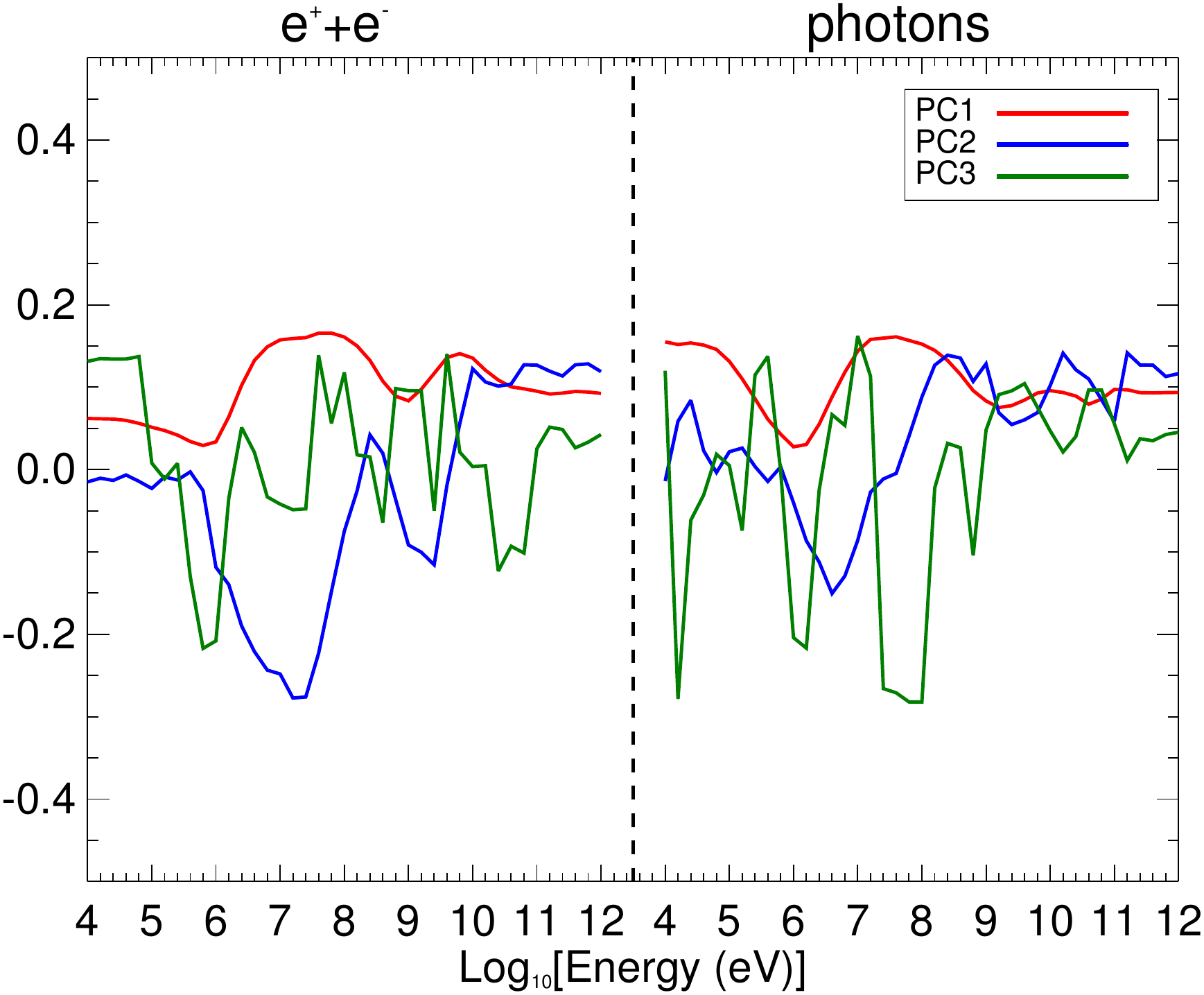}           
  \caption{
   The first three principal components for WMAP7 (left), \emph{Planck} (middle) and a CVL experiment (right), for annihilating DM (top), decaying DM with a long lifetime (middle) and decaying DM with a lifetime of $10^{13}$ s (bottom).}
\label{PCA}
\end{figure}

\section{Supplementary Materials}
\label{sec:supplementary}
We make available \texttt{.fits} and \texttt{.dat} files\footnote{http://nebel.rc.fas.harvard.edu/epsilon/} containing the values of the curves plotted in Figs.~\ref{fig:shortlifetimepca}-\ref{lifetimemcmc} . We also provide a Mathematica notebook to demonstrate the use of these files, with a worked example for how to compute the Planck constraints on the mass fraction of DM decaying to muons, for different lifetimes. 

There are two \texttt{.fits} files, each one with a corresponding \texttt{.dat} file.

\begin{itemize}
\item \textbf{energyPC1}:  This file contains the results plotted in Fig.~\ref{fig:shortlifetimepca}, containing arrays of the first PCs for annihilating DM and decaying DM with lifetimes $10^{13}, 10^{13.5}, 10^{14}, 10^{14.5}, 10^{15}$, and $10^{27}$ seconds. The first column (labeled ``log10energy'') gives the base 10 log of the (kinetic) energy in eV of one of the particles in the pair, the second column (labeled ``ann'') gives the result for annihilation, subsequent columns give the results for decay with lifetimes $10^{13}$ (``decay13''), $10^{13.5}$ (``decay135''), $10^{14}$ (``decay14''), $10^{14.5}$ (``decay145''), $10^{15}$ (``decay15''), and $10^{27}$ (``decay27'') seconds. The first 41 entries correspond to injection of $e^+e^-$ pairs, and the following 41 entries to injection of photons.
\item \textbf{lifetimePCA}:  The arrays in this file give the results of the PCA considering injection of 30 MeV $e^+e^-$ pairs and varying the decay lifetime, as plotted in Fig.~\ref{lifetimemcmc}. The first column (labeled ``log10lifetime'') gives the base 10 log of the decay lifetime in seconds. The second column (``PC1'') gives the Fisher-matrix forecast constraint using only the first PC, the third column (``PCsum'') gives the forecast constraint using the sum of the first five PCs, and the fourth column (``Normalized'') gives the forecast constraint based on the first five PCs, normalized so that for long lifetimes it matches the MCMC result.

\end{itemize}

\end{appendix}

\bibliography{decaypca}

\end{document}